\documentclass[fleqn,usenatbib]{mnras}
\usepackage[pagewise]{lineno}

\usepackage{newtxtext,newtxmath}

\usepackage[T1]{fontenc}

\DeclareRobustCommand{\VAN}[3]{#2}
\let\VANthebibliography\thebibliography
\def\thebibliography{\DeclareRobustCommand{\VAN}[3]{##3}\VANthebibliography}

\usepackage{graphicx}	
\usepackage{amsmath}	

\usepackage{xspace}

\newcommand\sfeff{\ensuremath{\epsilon_{\rm ff}}\xspace}   
\newcommand{\tform}{\ensuremath{t_\star}\xspace}  
\newcommand\Sigmacloud{\ensuremath{\Sigma_{\rm cloud}}\xspace} 
\newcommand\Mcl{\ensuremath{M_{\rm cl}}\xspace} 
\newcommand{\msun}{\ensuremath{M_{\odot}}\xspace}  
\newcommand{\avgNcomp}{\ensuremath{\langle N_{\mathrm{bound}} \rangle}\xspace}  
\newcommand{\Ncomp}{\ensuremath{N_{\mathrm{bound}}}\xspace} 
\newcommand{\avgNstar}{\ensuremath{\langle N_\star \rangle}\xspace}  
\newcommand{\ndensstar}{\ensuremath{n_{\rm S,b}}\xspace} 
\newcommand{\veldisstar}{\ensuremath{\sigma_{\rm S}}\xspace} 
\newcommand{\minmasscomp}{\ensuremath{M_{\mathrm{comp}}}\xspace} 
\newcommand{\STARFORGE}{\texttt{STARFORGE}\xspace} 
\newcommand{\finalstellarmass}{\ensuremath{M_{\rm stars}}\xspace}

\title[Stellar clustering around massive stars]{Star cluster formation from turbulent clumps. V.  Stellar clustering around massive stars}

\author[Gautam et al.]{
Aayush Gautam,$^{1,2}$\thanks{E-mail: gravitas908@gmail.com}
Juan P. Farias$^{3,4}$
and Jonathan C. Tan$^{1,2,5}$
\\
$^{1}$ Department of Astronomy, University of Virginia, Charlottesville, VA 22904, USA\\
$^{2}$ Virginia Institute for Theoretical Astronomy, University of Virginia, Charlottesville, VA 22904, USA \\
$^{3}$ Department of Astronomy, University of Texas at Austin, TX 78712, USA \\
$^{4}$ Department of Physics and Astronomy, McMaster University, 1280 Main Street West, Hamilton, ON, L8S 4M1, Canada \\
$^{5}$ Department of Physics \& Astronomy, Chalmers University of Technology, Gothenburg SE-41293, Sweden\\
}

\date{Accepted XXX. Received YYY; in original form ZZZ}

\pubyear{\the\year{}}

\begin{document}
\label{firstpage}
\pagerange{\pageref{firstpage}--\pageref{lastpage}}
\maketitle

\begin{abstract}
Massive stars ($\geq 8\:M_\odot$) are known to have high degrees of multiplicity, e.g., with about 60\% in triples or higher-order multiples. Such high levels of multiplicity may arise during formation (primary multiplicity) or through dynamical processing of already formed stars in dense clusters (secondary multiplicity). The level of primary multiplicity is an important metric to help distinguish between different formation scenarios, such as core accretion and competitive accretion. The level of secondary multiplicity is expected to evolve with time and be sensitive to local cluster environment. Here we analyze a suite of $N$-body simulations to study bound multiplicity and local projected stellar density, $N_\star$, around massive stars within gradually forming star clusters with 50\% primordial binaries in the Turbulent Clump Core Accretion (TCCA) paradigm. We find that massive stars rapidly gather triple or higher-order bound companions and enhancements in local $N_\star$ via dynamical processes. We study these metrics as a function of environment in a given cluster, quantifying the increasing multiplicity that arises towards cluster centers. We find that secondary multiplicity tends to decrease in more massive clusters due to their higher velocity dispersions, but rises as the mean density of the bound cluster increases. We find our $N_\star$ radial profiles are shallower compared to those in the STARFORGE simulations, which form massive stars via competitive accretion. A comparison to the AFGL 5180 system suggests it is better described by TCCA models. However, a larger number of observed systems is needed to better discriminate between these formation models.
\end{abstract}




\section{Introduction}
\label{sec:intro}

Massive stars, i.e., those with masses $\geq8\:M_\odot$, are both products and regulators of the process of star formation across the Universe. While short-lived and less numerous than their low-mass counterparts \citep{Kroupa2001MNRAS.322..231K, Chabrier2003PASP..115..763C}, massive stars strongly influence their star-forming environment, through protostellar outflows \citep{Tanaka2017ApJ...835...32T,KolliganKuiper2018A&A...620A.182K,Staff2019ApJ...882..123S,Staff2023ApJ...947...40S, Appel2026AJ....172...12A}, ionizing and dissociating radiation, and stellar winds and supernovae \citep[e.g.,][]{Matzner2002ApJ...566..302M, Dale2012MNRAS.424..377D, Dale2014MNRAS.442..694D,RogersPittard2013MNRAS.431.1337R,Butler2017ApJ...841...82B,Smith2018MNRAS.478..302S,Lewis2023ApJ...944..211L}.  Supernova explosions lead to the chemical enrichment of the interstellar medium through the dispersal of heavy elements synthesized inside massive stars \citep{Nomoto2013ARA&A..51..457N}, shaping the next generation of stars that will form in both isolated and clustered environments. Other feedback mechanisms affect star formation by restructuring the local gas distribution, possibly dispersing the parent cloud \citep{Lewis2023ApJ...944..211L, Appel2026AJ....172...12A}, however they could also trigger the formation of new stars during the process \citep[e.g.,][]{Klein1980SSRv...27..275K,Elmegreen2011EAS....51...45E,Kinoshita2021arXiv210805554K}. Considering their importance, the question of how massive stars form has been extensively studied both theoretically and observationally \citep[see, e.g.,][for reviews]{Tan2014prpl.conf..149T,Beuther2025ARA&A..63....1B}. 

The multiplicity of stars across the mass spectrum is an important constraint on formation theories. Observationally, the number of bound companions is known to rise with stellar mass \citep{Raghavan2010ApJS..190....1R,Sana2012Sci...337..444S,offner2023ASPC..534..275O}. In particular, massive stars have been found to have high degree of multiplicity across observational surveys: 90\% of stars above 8 \msun have at least one companion, and 55\% of them are part of triples and/or higher-order multiple systems \citep{MoeStefano2017ApJS..230...15M,offner2023ASPC..534..275O}. These higher-order multiple systems could have different configurations for their members and be hierarchical in nature: for example, a triple system could have an inner binary system with a wide tertiary \citep{Tokovinin2026arXiv260105006T, Tanikawa2026arXiv260121125T}; and a quadruple system could have two inner binaries bound to each other \citep{Zummer2025arXiv251115544Z}. These multiple systems could be formed through multiple pathways: filament fragmentation \citep{Inutsuka1997ApJ...480..681I,Pineda2015Natur.518..213P}, core fragmentation \citep{Kirk2017ApJ...838..114K,Guszejnov2017MNRAS.468.4093G}, and disc fragmentation \citep{Gammie2001ApJ...553..174G, Reynolds2021ApJ...907L..10R}. Remnant star-forming gas in the core or disc can lead to gas-driven migration of primordial companions during their formation \citep{BateBonnell1997MNRAS.285...33B,TokovininMoe2020MNRAS.491.5158T}. In addition, this primordial multiplicity is subject to further dynamical processing, with companions gathered or lost during dynamical interactions with other stars in the forming star cluster \citep{Heggie1975MNRAS.173..729H,Hills1975AJ.....80.1075H,Wall2019ApJ...887...62W}. 

The relative contribution of these mechanisms to the observed multiplicity remains an area of active research \citep{offner2023ASPC..534..275O}. These mechanisms should leave imprints on the separation and mass-ratio distribution of companions, however the completeness and bias of these observations should be carefully considered. Moreover, these primordial signatures could be masked by subsequent dynamical processing. So, it is important to study the multiplicity of massive stars in young star clusters before significant dynamical interactions have occurred \citep{offner2023ASPC..534..275O}. To complement this, one could use $N$-body simulations to quantify the level of dynamical processing that occurs on a population of primordial companions around massive stars. 

Different theories of massive star formation have different predictions about the primordial multiplicity of massive stars. For example, Turbulent Core Accretion \citep{McKee2003ApJ...585..850M}, especially in magnetically supported cores, can produce initially relatively isolated massive cores and thus stars. A certain level of clustering of companions around these massive stars could then develop subsequently via dynamical evolution in the environment of a forming star cluster. On the other hand, Competitive Accretion \citep{Bonnell2001MNRAS.323..785B, Wang2010ApJ...709...27W,Grudic2022MNRAS.512..216G} predicts the ubiquitous presence of many low-mass companions around a massive star, even from its earliest stages of formation. This presents a way to distinguish between these two massive star formation models by studying the clustering of lower-mass companions around massive stars in young star-forming regions. 

We expect that this may be accomplished by the consideration of surface density profiles of young stellar objects (YSOs) around massive protostars (i.e., while still accreting) and recently formed massive stars. Such surface number density profiles around massive stars have been studied using both observations \citep{CostaSilva2022A&A...659A..23C,Crowe2024A&A...682A...2C}, and simulations of star cluster formation \citep{Grudic2022MNRAS.512..216G}. Such comparisons with observed systems help us to understand whether massive stars can form in relative isolation or need to have companions present, shedding light on their dominant formation mechanism. 

An important complement to studies of primordial clustering around massive stars would be to investigate the effect of the star cluster environment on the ability of massive stars to capture companions through dynamical interactions. Such interactions are expected to be highly dependent on the local environment, particularly the local stellar velocity dispersion and stellar density. Both of these variables evolve with time and vary throughout the cluster, regulating the degree of gravitational focusing around massive stars, which would increase stellar captures. Furthermore, massive stars naturally segregate toward the centers of star clusters as they lose energy through interactions with lower-mass stars \citep{Allison2009ApJ...700L..99A,Dominguez2017MNRAS.472..465D,Chandrasekhar1943ApJ....97..255C}. Consequently, even within the same cluster, massive stars are expected to have greater chances to capture companions as they could move into the cluster center, where the local stellar density is much higher, and the dynamical environment is significantly different from the sparse outskirts. This effect may be even more pronounced if mass segregation is primordial \citep{Bonnell2001MNRAS.323..785B} rather than dynamical.

To capture all these processes, $N$-body simulations of gradually forming star clusters of different formation timescales are necessary. These simulations should also include accurate treatment of close encounters and regularization of close binary systems. In this paper, we use such existing $N$-body simulations of star cluster formation developed by \cite{Farias2019MNRAS.483.4999F}. In these simulations, star clusters form gradually within turbulent clumps in giant molecular clouds (GMCs) with a semi-analytic potential representing the gas that is slowly dissipated away. Massive stars are formed randomly in the star cluster, i.e., approximating the process of Turbulent Core Accretion in the limit of no primordial mass segregation of massive star formation. We perform an analysis of clustering of companions around these massive stars. We also compare with observations and massive stars formed by Competitive Accretion in \texttt{STARFORGE} simulations. In addition, the simulations involve 50\% of systems being primordial binaries, but with no preference for massive stars to be favored as being in multiples. There are no primordial triples or higher-order multiples, so such systems must form dynamically during the evolution of the cluster. We track the formation of such bound systems of companions around massive stars to study the effect of star cluster environment on the dynamical capture of companions. This exploration yields benchmark statistics that can be used to compare the results of more sophisticated MHD + $N$-body simulations to be developed in future studies.

\section{Methods}
\label{sec:methods}

In section~\ref{subsec:methods-starclusterformation}, we describe the $N$-body models of gradual star cluster formation used for this paper. We have post-processed these simulations to study the evolution of stellar clustering around massive stars. In section~\ref{subsec:methods-subsets-massivestars}, we describe the process of identifying bound systems around massive stars, which are later compared to study how the environment and formation timescale affect these bound systems. In section~\ref{subsec:methods-projectedstellardensities}, we describe how we construct projected stellar density profiles around massive stars to compare different models of massive star formation with each other and with observations. 

\subsection{Star Cluster Formation Simulations} 
\label{subsec:methods-starclusterformation}

In this work, we use the Turbulent Clump Core Accretion (TCCA) star cluster formation models developed by \cite{Farias2017ApJ...838..116F, Farias2019MNRAS.483.4999F,Farias2023MNRAS.523.2083F}, where star clusters are formed inside magnetized, turbulent, gravitationally bound, and initially starless clumps within giant molecular clouds (GMCs). These consist of a suite of unique $N$-body simulations focused on accurately following stellar dynamics in the early stages of star cluster formation, i.e., where there is still ongoing formation of stars. These models assume individual stars and primordial binary stars form via stochastic core accretion with initial clump properties based on the Turbulent Core Accretion (TCA) model \citep[][]{McKee2003ApJ...585..850M}.  
\cite{Farias2017ApJ...838..116F} (hereafter, Paper I) presented the extreme case of instantaneous star cluster formation, where all stars are formed together at once and remaining gas is assumed to leave the system instantaneously according to a given global star formation efficiency (SFE, $\epsilon$). \cite{Farias2019MNRAS.483.4999F} and \cite{Farias2023MNRAS.523.2083F} (hereafter, Papers II and III, respectively) presented the case of gradual star cluster formation. In these simulations, we assume that a fraction ($\epsilon$) of the gas is gradually converted into stars while the remaining gas is also steadily removed from the system due to stellar feedback. This is modeled as a linear decrease of the gas reservoir (Equation 9, Paper III) until it reaches zero at the end of star formation, \tform. The timescale over which this process occurs is set by the star formation efficiency per free-fall time, \sfeff, such that higher values of \sfeff result in shorter \tform (see Paper III for details). The case of \sfeff = $\infty$ corresponds to the instantaneous case where all stars are formed at once, whereas the slowest case considered assumes \sfeff = 0.01. For this paper, we post-processed clusters formed from $N$-body models outlined in Paper III.

In these pure $N$-body models, we emulate the parent cloud gas using an analytic potential that follows the structure from the TCCA model, i.e., a turbulent gas clump of mass \Mcl embedded in the surrounding molecular cloud. As mentioned above, the mass of this gas clump is linearly reduced as the star cluster forms, and the gas mass reaches zero at \tform. The parent cloud has a mass surface density, \Sigmacloud, which sets the bounding pressure of the clump: a clump of a given mass in a higher \Sigmacloud environment will be subject to higher pressure and thus be smaller and more dense. The characteristic radius of the clump ($R_{\rm cl}$) in a \Sigmacloud environment is:
\begin{eqnarray}
    R_{\rm cl} = 0.365 \left( \frac{\Mcl}{3000 M_\odot}\right)^{1/2} \left( \frac{\Sigmacloud}{1 \: \rm g\: cm^{-2}}\right)^{-1/2} \rm{pc}. 
    \label{eqn:clumpradius}
\end{eqnarray}
The clump is modeled as a polytropic sphere of gas with a three-dimensional density profile of the form:
\begin{eqnarray}
    \rho_{\rm cl} (r) = \rho_{\rm s,cl} \left( \frac{r}{R_{\rm cl}} \right)^{-k_{\rho}}.
    \label{eqn:polytropicdensity}
\end{eqnarray}
The density thus varies radially within the clump and is defined within the clump radius $R_{\rm cl}$. Here, $\rho_{\rm s,cl}$ is the density of clump at its edge ($r=R_{\rm cl}$). It is given by: 
\begin{eqnarray}
    \rho_{\rm s,cl} = \frac{(3-k_\rho)\Mcl}{4 \pi R_{\rm cl}^3}.
    \label{eqn:clumpsurfacedensity}
\end{eqnarray}
Similarly, the velocity dispersion profile of the clump is given by:
\begin{eqnarray}
    \sigma_{\rm cl} (r) = \sigma_{\rm s} \left( \frac{r}{R_{\rm cl}} \right)^{(2-k_{\rho})/2},
    \label{eqn:polytropicvelocitydispersion}
\end{eqnarray}
where $\sigma_{s}$ is the velocity dispersion at the surface of the clump, given by:
\begin{eqnarray}
    \sigma_{s} = 3.04 \left( \frac{\Mcl}{3000 M_\odot}\right)^{1/4} \left( \frac{\Sigmacloud}{1 \: \rm g\: cm^{-2}}\right)^{1/4} \rm{km \: s^{-1}}. 
    \label{eqn:clumpsurfacevelocitydispersion}
\end{eqnarray}
We set the exponent $k_{\rho}$ = 1.5 following Paper II, which are motivated by the observational results of \citet{Butler2012ApJ...754....5B}. The stars are formed following the above gas density (Equation~\ref{eqn:polytropicdensity}) and velocity dispersion (Equation~\ref{eqn:polytropicvelocitydispersion}) profiles of the parent clump. It is evident that clusters formed in higher \Mcl and \Sigmacloud models have larger stellar velocity dispersion $\sigma_{s}$. These simulations have no primordial mass segregation, however some degree of mass segregation is developed through dynamical processes, especially in the lower mass clusters (refer Section 4.4 of Paper III).

The time taken for the star cluster to form inside the clump is given by:
\begin{eqnarray}
    t_\star = \frac{\epsilon}{\sfeff} t_{\rm ff, 0},
    \label{eqn:formationtimescale}
\end{eqnarray}
where $t_{\rm ff,0}$ is the initial global freefall timescale of the clump,
\begin{eqnarray}
    t_{\rm ff,0} = 0.069 \left( \frac{\Mcl}{3000 M_\odot}\right)^{1/4} \left( \frac{\Sigmacloud}{1 \: \rm g\: cm^{-2}}\right)^{-3/4} \rm{Myr},
    \label{eqn:freefalltimescale}
\end{eqnarray}
in the TCCA model (Equation 9, Paper II). We find that for the same value of $\epsilon_{\rm ff}$, the timescale $t_\star$ is shorter for higher \Sigmacloud and/or smaller \Mcl, because the freefall timescale $t_{\rm ff,0}$ is shorter in higher \Sigmacloud and/or smaller \Mcl environments (see equation~\ref{eqn:freefalltimescale}). 

For our simulations, we fix $\epsilon$ to 0.5. For instance, for \Mcl = 3,000$\:\msun$ and $\epsilon$ = 0.5, we form 1,500 $\msun$ in stellar mass. We choose three different values of \Mcl: 300 \msun, 3,000 \msun and 30,000 \msun. For each \Mcl, we consider star clusters formed from clumps in two different cloud environments: set "L" formed from clumps in lower mass surface density clouds ($\Sigma_{\rm cloud}= 0.1\: \rm g\: cm^{-2}$), and set "H" formed from clumps in higher mass surface density clouds ($\Sigma_{\rm cloud}= 1.0\: \rm g\: cm^{-2}$). Next, for each \Sigmacloud, we consider five values for $\epsilon_{\rm ff}$: 0.01, 0.03, 0.10, 0.30, and 1.00. Table \ref{tab:simulations} shows the full set of cluster parameters (\Mcl, \Sigmacloud and \sfeff) modeled in this work. For each row of Table \ref{tab:simulations} we have run multiple realizations ($N_{\rm sims}$) shown in Column 2. The number of stars formed in a realization varies because of the random sampling of the IMF, so the average number of stars (and thus the average stellar mass) also differs slightly across models. However, on average, our models form around $\langle N_{\rm stars} \rangle \approx 4 \times 10^2, 4 \times 10^3$ and $4 \times 10^4$ stars in a realization of M300, M3000 and M30000 models, respectively. The resulting average stellar mass in our models is 0.375$\:\msun$. 

Each realization has an IMF sampling \citep{Kroupa2001MNRAS.322..231K} and 50\% primordial binary population as described in \cite{Farias2023MNRAS.523.2083F}. However, we do not have primordial triples, quadruples or higher-order multiples. Any systems with three or more bound stars must have formed via dynamical processing. We track the formation of such systems around massive stars studied for this work. To quantify the formation of such bound systems, we calculate the fraction of massive stars that have triple or higher-order companions around them. This fraction called the Triple/Higher-order fraction (THF) is given by,
\begin{eqnarray}
    \rm THF = \frac{T+Q+P+\cdots}{S+B+T+Q+P+\cdots},
    \label{eqn:thf}
\end{eqnarray}
where $\rm S, B, T, Q, P$ are the number of single stars, binaries, triples, quadruples, pentuples, and so on \citep{offner2023ASPC..534..275O}.  

\begin{table*}
\centering
\caption{Overview of star cluster formation simulations. First column is the name of the set. Second column is number of simulations ($N_{\rm sims}$) run in each set. Third column is the star formation efficiency per freefall time (\sfeff) for the set. Fourth column is the mass surface density of the parent cloud (\Sigmacloud). Column 5 is the mass of the gas clump (\Mcl). Sixth column is the number of stars ($\langle N_{\rm stars} \rangle$) formed on average in the model. Columns 7 and 8 are the star formation time and initial free-fall time of the clump, respectively. Column 9 is the radius of the gas clump ($R_{\rm cl}$). Column 10 is the velocity dispersion at the surface of the clump.}

\label{tab:simulations}
\begin{tabular}{rccccccccc}
Set Name & $N_{\rm sims}$ & \sfeff & \Sigmacloud & \Mcl & $\langle N_{\rm stars} \rangle$ & $t_{\star}$ & $t_{\rm ff,0}$ & $R_{\rm cl}$ & $\sigma_{s}$ \\ 
& & & [$\rm g\:cm^{-2}$] & [$M_\odot$] &  & [Myr] & [Myr] & [pc] & [$\rm km\: s^{-1}$] \\ \hline
  & 200 & 0.01  & 0.1 & 300    & $4 \times 10^2$   & 10.91 & 0.22  & 0.36 & 0.96 \\
  & 200 & 0.03  & 0.1 & 300    & $4 \times 10^2$    & 3.64  & 0.22  & 0.36 & 0.96 \\
M300L  & 200 & 0.1   & 0.1 & 300    & $4 \times 10^2$    & 1.09  & 0.22  & 0.36 & 0.96 \\
  & 200 &  0.3   & 0.1 & 300    & $4 \times 10^2$   & 0.36  & 0.22  & 0.36 & 0.96 \\
  & 200 & 1.0   & 0.1 & 300    & $4 \times 10^2$    & 0.11  & 0.22  & 0.36 & 0.96 \\\cline{2-10}

 & 200 & 0.01   & 1.0 & 300    & $4 \times 10^2$   & 1.94  & 0.039  & 0.115 & 1.71 \\
 & 200 & 0.03   & 1.0 & 300    & $4 \times 10^2$   & 0.65  & 0.039  & 0.115 & 1.71 \\
M300H & 200 &  0.1    & 1.0 & 300    & $4 \times 10^2$    & 0.19  & 0.039  & 0.115 & 1.71 \\
 & 200 & 0.3    & 1.0 & 300    & $4 \times 10^2$    & 0.06  & 0.039  & 0.115 & 1.71 \\
 & 200 & 1.0    & 1.0 & 300    & $4 \times 10^2$    & 0.02  & 0.039  & 0.115 & 1.71 \\\cline{1-10}
  
  & 20 & 0.01 & 0.1 & 3,000 & $4 \times 10^3$  & 19.40 & 0.39  & 1.15 & 1.71 \\
  & 20 & 0.03 & 0.1 & 3,000 & $4 \times 10^3$  & 6.47  & 0.39  & 1.15 & 1.71 \\
M3000L  & 20 & 0.1 & 0.1 & 3,000 & $4 \times 10^3$  & 1.94  & 0.39  & 1.15 & 1.71 \\
  & 20 & 0.3  & 0.1 & 3,000 & $4 \times 10^3$  & 0.65  & 0.39  & 1.15 & 1.71 \\
  & 20 & 1.0  & 0.1 & 3,000 & $4 \times 10^3$  & 0.19  & 0.39  & 1.15 & 1.71 \\ \cline{2-10}
  
 &  20 & 0.01 & 1.0 & 3,000 & $4 \times 10^3$  & 3.45  & 0.069  & 0.365 & 3.04 \\
 &  20 & 0.03 & 1.0 & 3,000 & $4 \times 10^3$  & 1.15  & 0.069  & 0.365 & 3.04 \\
M3000H & 20 & 0.1  & 1.0 & 3,000 & $4 \times 10^3$  & 0.35  & 0.069  & 0.365 & 3.04 \\
 & 20 & 0.3  & 1.0 & 3,000 & $4 \times 10^3$ & 0.12  & 0.069  & 0.365 & 3.04 \\
 & 20 & 1.0  & 1.0 & 3,000 & $4 \times 10^3$  & 0.03  & 0.069  & 0.365 & 3.04 \\ \cline{1-10}

  & 2 & 0.01 & 0.1 & 30,000 & $4 \times 10^4$ & 34.50 & 0.69  & 3.65 & 3.04 \\
  & 2 & 0.03 & 0.1 & 30,000 & $4 \times 10^4$ & 11.50 & 0.69  & 3.65 & 3.04 \\
 M30000L & 2 &  0.1  & 0.1 & 30,000 & $4 \times 10^4$ & 3.45 & 0.69  & 3.65 & 3.04 \\
  & 2 & 0.3  & 0.1 & 30,000 & $4 \times 10^4$ & 1.15 & 0.69  & 3.65 & 3.04 \\
  & 2 & 1.0  & 0.1 & 30,000 & $4 \times 10^4$ & 0.34 & 0.69  & 3.65 & 3.04 \\ \cline{2-10}

 & 2 & 0.01 & 1.0 & 30,000 & $4 \times 10^4$ & 6.14  & 0.123  & 1.154 & 5.41 \\
 & 2 & 0.03 & 1.0 & 30,000 & $4 \times 10^4$& 2.05  & 0.123  & 1.154 & 5.41 \\
 M30000H & 2 &  0.1  & 1.0 & 30,000 & $4 \times 10^4$ & 0.61  & 0.123  & 1.154 & 5.41 \\
 & 2 & 0.3  & 1.0 & 30,000 & $4 \times 10^4$ & 0.20  & 0.123  & 1.154 & 5.41 \\
 & 2 & 1.0  & 1.0 & 30,000 & $4 \times 10^4$ & 0.06  & 0.123  & 1.154 & 5.41 \\ \cline{1-10}
\end{tabular}
\end{table*}

\subsection{Clustering around Massive Stars}
\label{subsec:methods-subsets-massivestars}

We examine the stellar clustering around massive stars ($\geq 8\: \msun$) formed in our simulations. To study the effect of cluster environment on such clustering, we divide these massive stars into three different subsets according to their location and boundedness. First, we determine stars that are bound to the star cluster using a snowballing algorithm developed in \cite{Farias2023MNRAS.523.2083F}. Then, we refer bound massive stars in the central regions within 20\% of the half-mass radius of their star cluster as \textit{central stars}. Similarly, massive stars beyond 20\% of the half-mass radius which are in the outer regions of the star cluster are referred as \textit{offset stars}. Finally, massive stars that are unbound from the star cluster are referred to as \textit{ejected stars}. 

After dividing massive stars into subsets, we study the properties of bound and unbound companions around them. For this purpose, we have developed routines to identify bound companions around any given massive star. The routine works as follows: First, we identify the closest bound companion (a binary partner) to the massive star. Then, we replace the two stars with a center of mass particle. Next, we search for the next closest star that is bounded to this center of mass particle, hence forming a triple. If present, we replace the triple with its corresponding center of mass particle. Then, we search for the next closest bound companion. We perform this process iteratively to identify all the stars that are bound to the massive star. We start this search around the most massive star, identify its bound companions and tag them. Then, we move to the next most massive untagged star and search for its bound companions, ensuring there is no double-counting of stars during this iterative process. This allows us to identify all the bound stars that are centered on independent primary massive stars within our star cluster. 

At each step, the bound systems are identified based on the total kinetic and potential energies of the components, combined with a tidal force criterion (equation \ref{eqn:tidalforce_criterion}) to remove unstable companions. If $m_S$ and $m_\star$ are the masses of the two particles under consideration separated by distance $d_1$, and $m_p$ is the mass of a perturber star at a distance $d_2$ from $m_\star$, we are checking if $m_\star$ is bound to $m_S$ under perturbation from a perturber star $m_p$:  
\begin{eqnarray}
    \left| \frac{m_S m_\star}{d_1^2} \right| > \left| K \frac{m_p m_\star}{d_2^2} \right|.
    \label{eqn:tidalforce_criterion}
\end{eqnarray}
Here $m_S$ could be a single star or a center of mass particle in the case of an inner system of bound stars. We set $K$ = 10. To calculate the internal and external forces, we iterate and sum up contributions from all stars internal and external to the bound system, respectively.  

We denote \Ncomp as the total number of bound companions for a given massive star. We study any correlations between \Ncomp and the mass of the massive star $M_\star$, and its radial location in the star cluster. We define the radial location of the star in its star cluster using a radial coordinate $r[r_h] = r_\star/r_h$, where we normalize the radial distance of the star ($r_\star$) from the center by the half-mass radius ($r_h$) of the bound cluster. The center of the bound cluster is computed using the snowballing algorithm of \cite{Farias2023MNRAS.523.2083F} mentioned above.  

\subsection{Projected stellar density profiles}
\label{subsec:methods-projectedstellardensities}

After the identification of bound and unbound stars around every massive star, we construct bound and total (bound+unbound) projected stellar density profiles for the massive star. For this, we create annular bins centered on the massive star. For the linear case, we create annular bins of width 0.05 pc centered on the massive star - the innermost bin is a circle of radius 0.05 pc. For logarithmic case, we choose $N_{\rm bins}$ 30 bins between $10^{-5}$ pc and 1 pc to probe the small-scale distribution near the massive star. The distances between the stars are 2D separations measured in projection along the $z-$axis in the simulation. Then, we count the number of stars $n_\star (r)$ in each radial bin of area $A(r)$ located at radius $r$ from the massive star.  We convert this into a projected 2D stellar density $N_\star$:
\begin{eqnarray}
    N_\star (r) = \frac{n_\star (r)}{A(r)}. 
    \label{eqn:Nstar}
\end{eqnarray}

Here, $n_\star = n_{\rm bound} + n_{\rm unbound}$, where $n_{\rm bound}$ is the number of bound companions and $n_{\rm unbound}$ is the number of unbound companions in a given radial bin. Adding up bound companions in all bins, we recover \Ncomp. Using equation~\ref{eqn:Nstar}, we construct the bound and total projected stellar density profiles around every massive star. We compare these profiles across our full range of models and different cluster ages. We also compare them with observational data to explore our questions about theoretical models of massive star formation.   

\section{Results}
\label{sec:results}

In section~\ref{subsec:results-location&mass}, we present our results on the variation of the number of bound companions around a massive star with its location and mass. In section~\ref{subsec:results-modelcomparison}, we compare massive stars across star clusters forming at different timescales in our (\Mcl,\Sigmacloud,\sfeff) parameter space. In section~\ref{subsec:results-projectedstellardensities}, we present the evolution of projected stellar density profiles around these massive stars. In section~\ref{subsec:results-observations}, we compare these profiles and the properties of bound companions with observations.  

Before we begin, we note that typical, well-studied young star-forming regions (ages $\lesssim$ 10 Myr) have a wide range of ages: 1-2 Myr \citep[e.g., Ophiuchus;][]{Wilking2008hsf2.book..351W}, 2-5 Myr \citep[e.g., Taurus, Lupus, Chamaeleon I;][]{Kenyon1995ApJS..101..117K, Alcala2014A&A...561A...2A,Luhman2004ApJ...602..816L} and 5-10 Myr \citep[e.g., Upper Scorpius;][]{Pecaut2012ApJ...746..154P}. One of the most notable is the Orion Nebula Cluster with stellar ages between $\sim$1 and $\sim$5 Myr \citep[e.g.,][]{DaRio2010ApJ...722.1092D,Reggiani2011A&A...534A..83R,Fajrin2025MNRAS.537.1320F}. These age estimates have their own systematic uncertainties \citep[e.g.,][]{Preibisch2012RAA....12....1P}, which are beyond the scope of our work. So, we have picked 3 Myr as a fiducial age for a typical young star-forming region. This adopted reference time can influence the trends we observe during our model comparisons at a given cluster age. Our different star clusters will have reached different stages at 3 Myr in terms of how advanced the star formation process is. However, this is also true for the observed systems; and unlike the simulations, we do not know which exact stage of the process a given star-forming region is currently in. So, in order to make a fair comparison between theoretical models in this work, we need to pick a fixed physical time (3 Myr). When we compare models with observations as in section~\ref{subsec:results-observations}, we generate the metrics at different ages to account for the uncertainty in our assumption for the typical age of the star-forming region. 

\begin{table*}
\centering
\caption{Statistics of primary massive stars ($M_{\rm prim} \geq 8\: M_\odot$) categorized into \textit{central}, \textit{offset} and \textit{ejected} subsets in our star cluster formation simulations. Primary massive stars are massive stars that act as primaries for their own independent bound stellar systems, so there is no double counting of stars. The statistics presented in this table were calculated at a cluster age $t$ = 3.0 Myr. Since these are models of gradual star cluster formation (varying \sfeff, column 2), some models are still forming stars (including massive stars) while others have fully formed their star clusters by this age, leading to a variation in $N_{\rm prim}$ at $t$ = 3.0 Myr (column 3). Columns 4, 5 and 6 show the relative number of \textit{central}, \textit{offset} and \textit{ejected} stars. Columns 7 to 9 shows the number of \textit{central} stars that are singles (S: \Ncomp = 0), in binaries (B: \Ncomp = 1), or part of triple/higher-order bound stellar systems (TH: $\Ncomp \geq 2$). Columns 10 to 12 show the same for \textit{offset} stars, and columns 13 to 15 for \textit{ejected} stars, respectively.}

\label{tab:massivestars}
\begin{tabular}{rcccccccccccccc}
Set Name & $\sfeff$ & $N_{\rm prim}$ & $N_{\rm prim}$ & $N_{\rm prim}$ & $N_{\rm prim}$ & \multicolumn{3}{c}{$N_{\rm prim, central}$} & \multicolumn{3}{c}{$N_{\rm prim, offset}$} & \multicolumn{3}{c}{$N_{\rm prim, ejected}$} \\ 
& & $(M_{\rm prim} \geq 8\: M_\odot)$ & (central) & (offset) & (ejected) & S & B & TH & S & B & TH  & S & B & TH \\ \hline
  & 0.01 & 204 & 69 & 134 & 1 & 19 & 28 & 22 & 58 & 58 & 18 & 1 & 0 & 0 \\ 
  & 0.03 & 214 & 128 & 73 & 13 & 39 & 33 & 56 & 27 & 34 & 12 & 7 & 6 & 0 \\ 
M300L & 0.10 & 255 & 161 & 76 & 18 & 31 & 44 & 86 & 29 & 29 & 18 & 13 & 3 & 2 \\ 
  & 0.30 & 248 & 138 & 75 & 35 & 28 & 38 & 72 & 24 & 27 & 24 & 17 & 16 & 2 \\ 
  & 1.00 & 268 & 134 & 88 & 46 & 33 & 37 & 64 & 22 & 27 & 39 & 21 & 19 & 6 \\\cline{2-15}
  & 0.01 & 276 & 166 & 75 & 35 & 23 & 43 & 100 & 33 & 19 & 23 & 22 & 11 & 2 \\ 
  & 0.03 & 190 & 122 & 42 & 26 & 17 & 29 & 76 & 12 & 13 & 17 & 16 & 7 & 3 \\ 
M300H & 0.10 & 243 & 146 & 56 & 41 & 20 & 36 & 90 & 12 & 16 & 28 & 24 & 14 & 3 \\ 
  & 0.30 & 256 & 157 & 52 & 47 & 26 & 41 & 90 & 6 & 12 & 34 & 17 & 20 & 10 \\ 
  & 1.00 & 270 & 132 & 67 & 71 & 15 & 34 & 83 & 10 & 15 & 42 & 30 & 23 & 18 \\\cline{1-15}
  & 0.01 & 57 & 5 & 47 & 5 & 3 & 2 & 0 & 34 & 12 & 1 & 4 & 1 & 0 \\ 
  & 0.03 & 140 & 11 & 122 & 7 & 4 & 3 & 4 & 69 & 50 & 3 & 4 & 3 & 0 \\ 
M3000L & 0.10 & 251 & 39 & 170 & 42 & 11 & 14 & 14 & 95 & 69 & 6 & 21 & 20 & 1 \\ 
  & 0.30 & 264 & 64 & 133 & 67 & 31 & 25 & 8 & 77 & 44 & 12 & 40 & 25 & 2 \\ 
  & 1.00 & 259 & 61 & 121 & 77 & 23 & 27 & 11 & 68 & 40 & 13 & 44 & 30 & 3 \\\cline{2-15}
  & 0.01 & 197 & 48 & 136 & 13 & 14 & 25 & 9 & 81 & 50 & 5 & 8 & 5 & 0 \\ 
  & 0.03 & 278 & 63 & 139 & 76 & 31 & 23 & 9 & 98 & 36 & 5 & 53 & 23 & 0 \\ 
M3000H & 0.10 & 258 & 60 & 118 & 80 & 26 & 24 & 10 & 73 & 35 & 10 & 50 & 25 & 5 \\ 
  & 0.30 & 254 & 71 & 106 & 77 & 32 & 24 & 15 & 51 & 40 & 15 & 48 & 23 & 6 \\ 
  & 1.00 & 261 & 63 & 110 & 88 & 26 & 20 & 17 & 61 & 30 & 19 & 52 & 31 & 5 \\\cline{1-15}
  & 0.01 & 20 & 3 & 17 & 0 & 2 & 1 & 0 & 11 & 6 & 0 & 0 & 0 & 0 \\ 
  & 0.03 & 68 & 9 & 57 & 2 & 5 & 4 & 0 & 35 & 22 & 0 & 2 & 0 & 0 \\ 
M30000L & 0.10 & 225 & 19 & 185 & 21 & 12 & 7 & 0 & 114 & 70 & 1 & 13 & 8 & 0 \\ 
  & 0.30 & 259 & 37 & 167 & 55 & 22 & 11 & 4 & 101 & 62 & 4 & 37 & 18 & 0 \\ 
  & 1.00 & 255 & 34 & 160 & 61 & 20 & 13 & 1 & 95 & 62 & 3 & 32 & 28 & 1 \\\cline{2-15}
  & 0.01 & 142 & 10 & 130 & 2 & 7 & 1 & 2 & 89 & 40 & 1 & 2 & 0 & 0 \\ 
  & 0.03 & 268 & 43 & 181 & 44 & 25 & 17 & 1 & 109 & 71 & 1 & 31 & 13 & 0 \\ 
M30000H & 0.10 & 260 & 38 & 174 & 48 & 24 & 13 & 1 & 112 & 58 & 4 & 27 & 21 & 0 \\ 
  & 0.30 & 262 & 55 & 142 & 65 & 34 & 19 & 2 & 90 & 51 & 1 & 39 & 26 & 0 \\ 
  & 1.00 & 273 & 38 & 169 & 66 & 28 & 8 & 2 & 116 & 49 & 4 & 49 & 17 & 0 \\\cline{1-15}
\end{tabular}
\end{table*}

\subsection{Effect of location and stellar mass}
\label{subsec:results-location&mass}

We classified massive stars into three subsets: \textit{central stars}, \textit{offset stars} and \textit{ejected stars}
to study the effect of cluster environment on their bound companions. Table~\ref{tab:massivestars} presents statistics of primary massive stars across these subsets in our simulations across all (\Mcl, \Sigmacloud, \sfeff) models at $t$ = 3 Myr. Figures below show the results for our fiducial \sfeff = 0.03 model, unless otherwise specified. These primary massive stars have their own independent systems of bound companions (if any). We find that \textit{ejected} subset comprised a range of 1-30\% of the primary massive star population at $t$ = 3.0 Myr, depending on the model. The dynamical state of the forming star cluster determined the relative fraction of \textit{central} and \textit{offset} stars in the remaining stars that are still bound to the cluster. 

\begin{figure*}
    \centering
    \includegraphics[width=\linewidth]{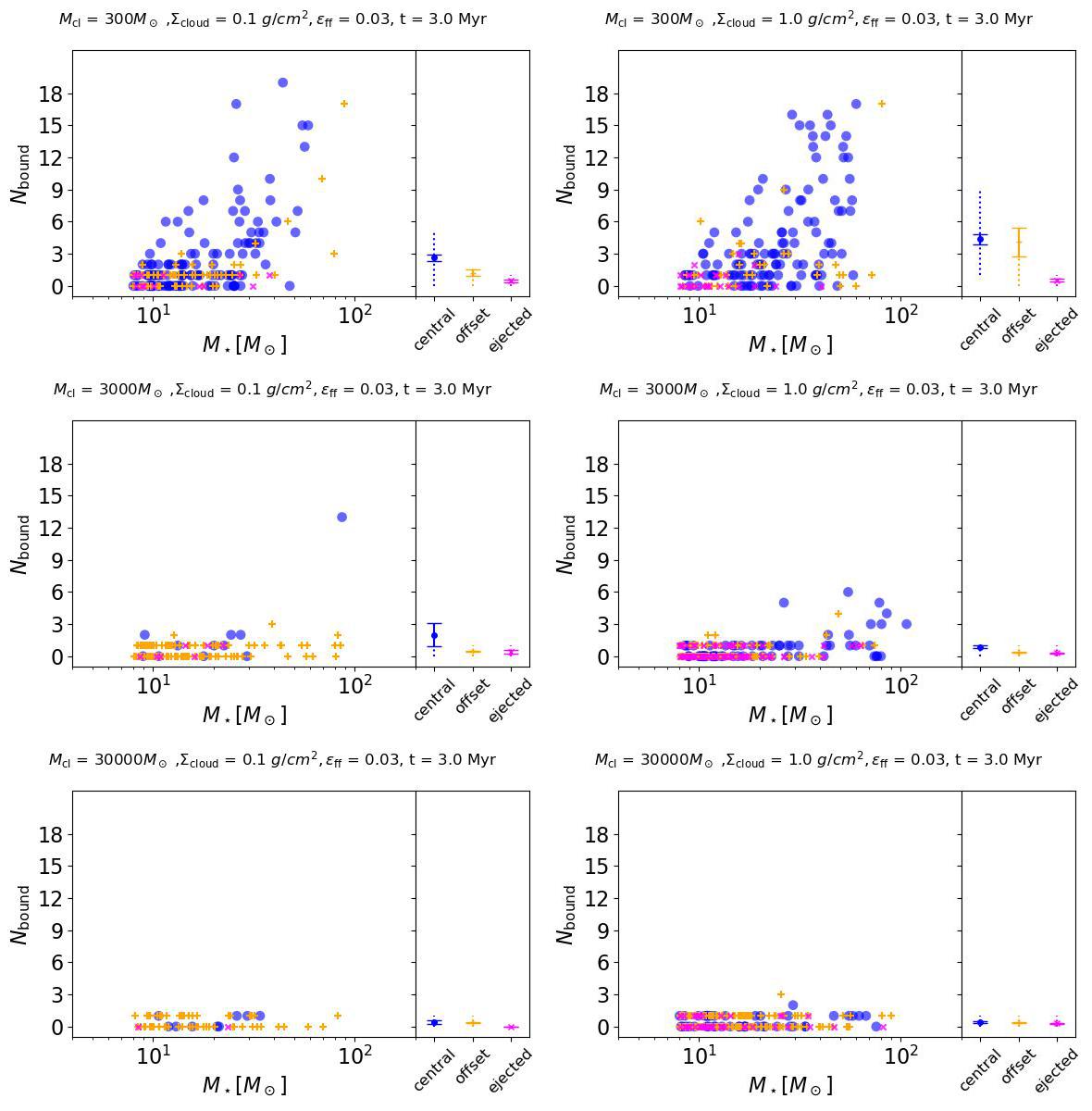}
    \caption{Masses of massive stars ($M_\star$) versus the number of bound companions (\Ncomp) around them, in our fiducial $\sfeff = 0.03$ models at cluster age $t = $ 3.0 Myr. From top to bottom, the rows show M300, M3000, and M30000 models, respectively. The left and right columns show low ($\Sigmacloud = 0.1\: \rm g\: cm^{-2}$) and high ($\Sigma_{\rm cloud}= 1.0\: \rm g\: cm^{-2}$) cloud mass surface density cases, respectively. In each panel, the colored symbols (blue circles, orange pluses, magenta crosses) show the three subsets of massive stars: \textit{central} stars, \textit{offset} stars and \textit{ejected} stars, respectively. In each side panel, we show the mean number of companions, \avgNcomp,  averaged around massive stars in each subset. The dotted errorbar indicates the spread of \Ncomp from the 16th percentile to the 84th percentile, while the solid errorbar shows the standard error in \avgNcomp. We find that more massive stars have a larger number of bound companions, however this also depends on their location within the star cluster.}
    \label{fig:massivestar_mass_ncompanions}
\end{figure*}

In Figure~\ref{fig:massivestar_mass_ncompanions}, we show the mass $M_\star$ of the massive stars versus the number of bound companions ($N_{\rm comp}$) around them. We find that more massive stars tend to have more companions than lower-mass stars, as expected. However, \Ncomp depends on the environment of the massive star as well. For similar $M_\star$, two massive stars can have different \Ncomp in the same star cluster depending on their location. We find that a \textit{central star} tends to have more companions than an \textit{ejected star} of the same mass. For instance, we find that \textit{ejected stars} tend to be mostly singles (\Ncomp = 0) or binaries (\Ncomp = 1) across all our models. Only a few of them have more than one companion, with those companions having small mass-ratios compared to the massive star. 

\begin{figure*}
    \centering
    \includegraphics[width=\linewidth]{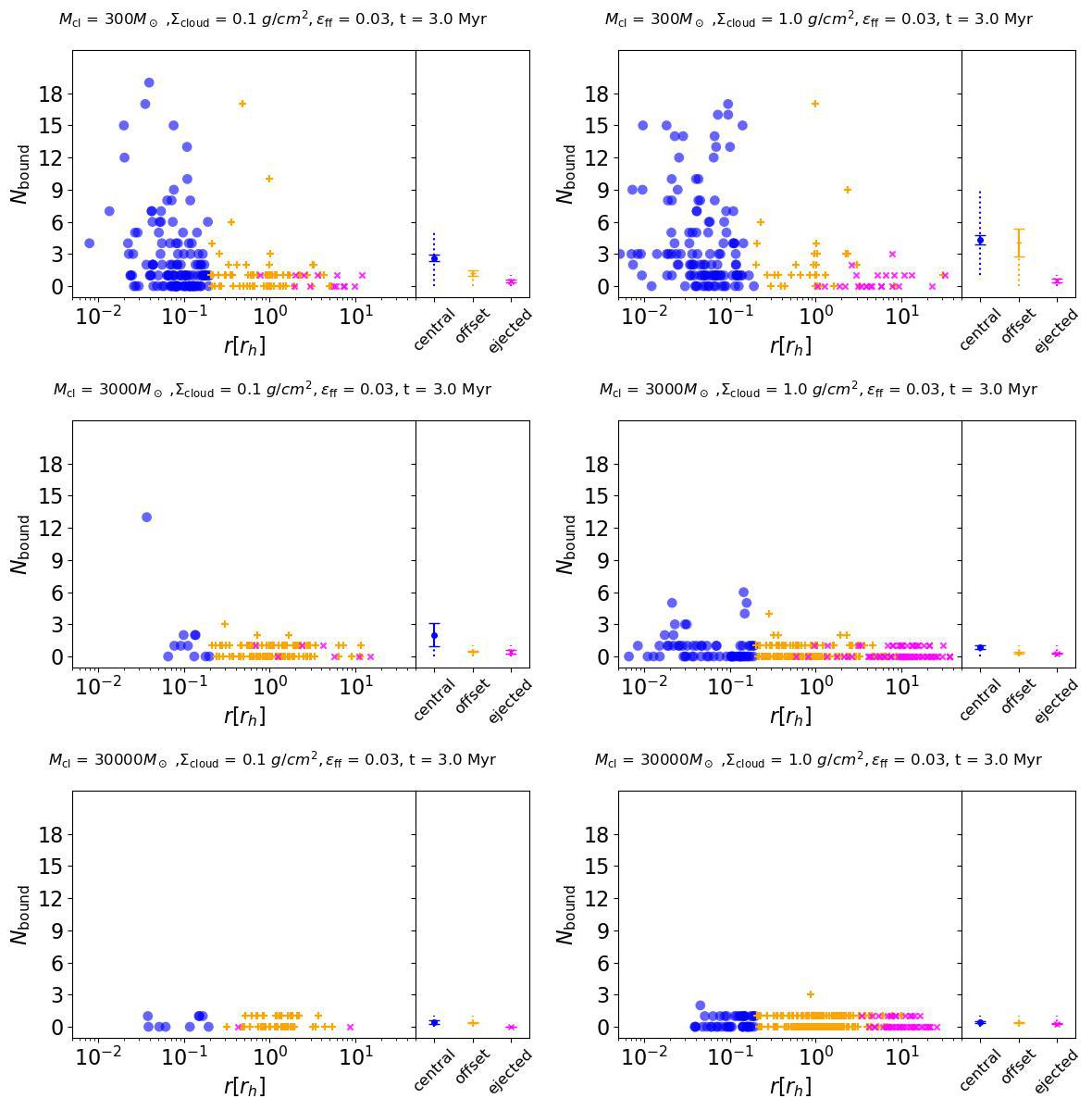}
    \caption{Normalized radial locations of massive stars ($r_h$) versus number of bound companions (\Ncomp) around them, in our fiducial $\sfeff = 0.03$ models at cluster age $t = $ 3.0 Myr. From top to bottom, the rows show M300, M3000, and M30000 models, respectively. The left and right columns show low ($\Sigmacloud = 0.1\: \rm g\: cm^{-2}$) and high ($\Sigma_{\rm cloud}= 1.0\: \rm g\: cm^{-2}$) cloud mass surface density cases, respectively. In each panel, the colored symbols (blue circles, orange pluses, magenta crosses) show three subsets of massive stars: \textit{central} stars, \textit{offset} stars and \textit{ejected} stars, respectively. In each side panel, we show the mean number of companions, \avgNcomp, averaged around massive stars in each subset. The dotted errorbar indicates the spread of \Ncomp from the 16th percentile to the 84th percentile, while the solid errorbar shows the standard error in \avgNcomp. We find that massive stars which are closer to the center of the star cluster generally have more bound companions.}
    \label{fig:massivestar_location_ncompanions}
\end{figure*}

To further illustrate these effects, in Figure~\ref{fig:massivestar_location_ncompanions}, we show $N_{\rm comp}$ versus the normalized radial location ($r[r_h]$) of the massive stars. First, we can infer the distribution of the three subsets of massive stars inside their star clusters. For instance, the \textit{ejected stars} are located mostly beyond 0.5 times the half-mass radius of their cluster. Similarly, we observe that many massive stars ($\geq 20 \msun$) have segregated to the center of their star cluster ($r[r_h] \leq 0.2$). Next, we find that there is some correlation between \Ncomp and $r[r_h]$. \textit{Central stars} have more companions in general than \textit{offset stars}. Also, some \textit{central} stars deeper in the star cluster have more companions than \textit{central} stars of similar mass. In addition to the variation of \Ncomp between massive stars in the same star cluster, there are clear variations in \Ncomp across star clusters in different models (compare different panels of Figure~\ref{fig:massivestar_mass_ncompanions} and \ref{fig:massivestar_location_ncompanions}) of our parameter space (\Mcl, \Sigmacloud, \sfeff). We discuss these differences in section~\ref{subsec:results-modelcomparison}. 

\subsection{Comparison across models}
\label{subsec:results-modelcomparison}

In order to understand the effect of the environment of a forming star cluster on stellar clustering around massive stars, we compared the statistics of bound companions around these massive stars across our full range of models (\Mcl,\Sigmacloud,\sfeff) outlined in Table~\ref{tab:simulations}. For this, we averaged the number of bound companions \Ncomp across all massive stars in each subset to find the average number of bound companions (\avgNcomp). Then, we studied the variation of \avgNcomp with our model parameters (\Mcl, \Sigmacloud and \sfeff). 

\begin{figure*}
    \centering
    \includegraphics[width=\textwidth]{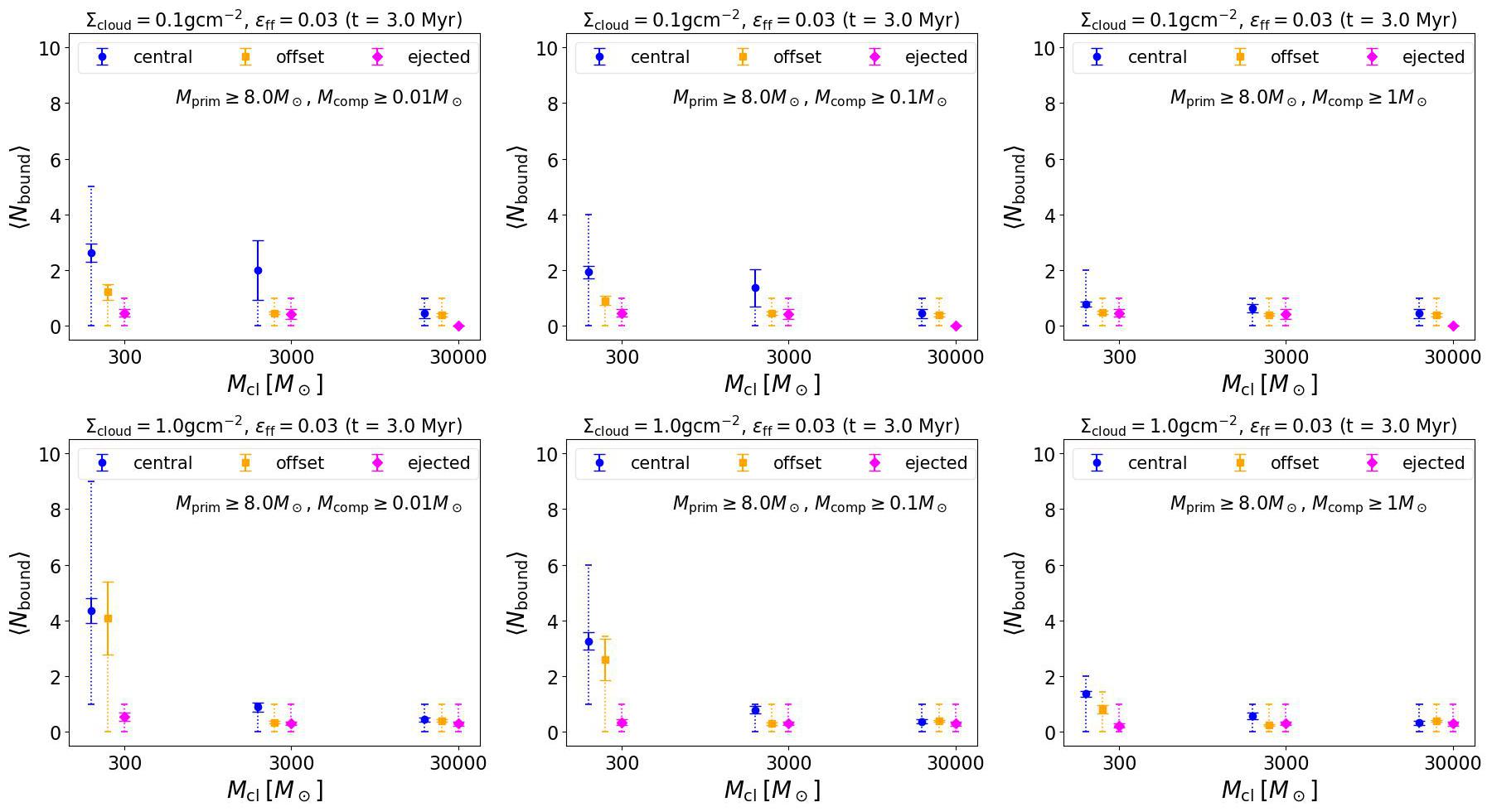}
    \caption{Each panel in the grid shows the comparison between different \Mcl models showing the average number of bound companions \avgNcomp around massive stars in that model. Panels in the first and second rows show the low ($\Sigmacloud = 0.1\: \rm g\: cm^{-2}$) and high ($\Sigmacloud = 1.0\: \rm g\: cm^{-2}$) cloud mass surface density cases, respectively, for the fiducial (\sfeff = 0.03) models at cluster age of $t = 3.0 $ Myr. In the columns, we vary the threshold for companion mass (\minmasscomp) between 0.01 \msun (fiducial), 0.1 \msun and 1 \msun. Each panel, on the $y-$axis, shows \avgNcomp around massive stars in three subsets: \textit{central} stars (blue), \textit{offset} stars (orange) and \textit{ejected} stars (magenta). $\langle ... \rangle$ indicates averaging over massive stars in all realizations of each model. The points for different subsets are offset by [0.1,0.0,-0.1] along $x-$axis to allow for visual comparison. For each point, the dotted errorbar indicates the spread of \Ncomp from the 16th percentile to the 84th percentile, while the solid errorbar shows the standard error in \avgNcomp. We find that massive stars in smaller \Mcl star clusters with lower velocity dispersion have more bound companions.}
\label{fig:model_comparison_variation_Mprim_Mcomp_fullset}
\end{figure*}

Figure~\ref{fig:model_comparison_variation_Mprim_Mcomp_fullset} shows \avgNcomp around massive stars in different \Mcl models at a cluster age of $t = 3.0$ Myr. In addition to \textit{central} stars having more companions than \textit{ejected} stars for all \Mcl models, we find that massive stars in the smaller \Mcl clusters have larger \avgNcomp compared to those in higher-mass clusters. This is attributed to the lower velocity dispersion (\veldisstar) for smaller \Mcl clusters in our TCM models (refer column 10 of Table~\ref{tab:simulations}). Finally, the \textit{central} massive stars in higher \Sigmacloud models have slightly more companions than lower \Sigmacloud models. This is due to the smaller half-mass radius of these higher \Sigmacloud models (refer column 9 of Table~\ref{tab:simulations}), which leads to higher stellar densities (\ndensstar) in the bound cluster. We investigate the variation of \avgNcomp with \veldisstar and \ndensstar later in this section. 

We studied the potential effect of observational limits on the statistics of \avgNcomp. These limits mean only companions above a certain mass would be observable. For this, we considered three cases for companion masses: $\minmasscomp \geq 0.01 \msun $ (fiducial case), $\minmasscomp \geq 0.1 \msun$ and $\minmasscomp \geq 1.0 \msun$. Here, 0.01~\msun is the lowest possible mass of objects in our star cluster simulations and 0.1~\msun is the observational limit on companion mass from the AFGL 5180 observations, which we later compare with our simulated companions. We chose 1.0~\msun to follow a uniform logarithmic order in companion mass. Figure~\ref{fig:model_comparison_variation_Mprim_Mcomp_fullset} shows the variation of \avgNcomp with these choices for the low and high \Sigmacloud models in top and bottom rows, respectively. We find that if we consider only higher-mass companions by increasing the companion mass threshold ($M_{\rm comp}$), the metric \avgNcomp goes down as expected. The decline is more prominent for the lower \Mcl models, which indicates massive stars in these models have many low-mass companions below 0.1 \msun. In addition, the \avgNcomp around \textit{ejected} stars is not as affected as much as \textit{central} or \textit{offset} stars. This indicates that these \textit{ejected} stars usually have higher-mass companions ($> 1.0 \msun$), if any. 

\begin{figure*}
    \centering
    \includegraphics[width=\textwidth]{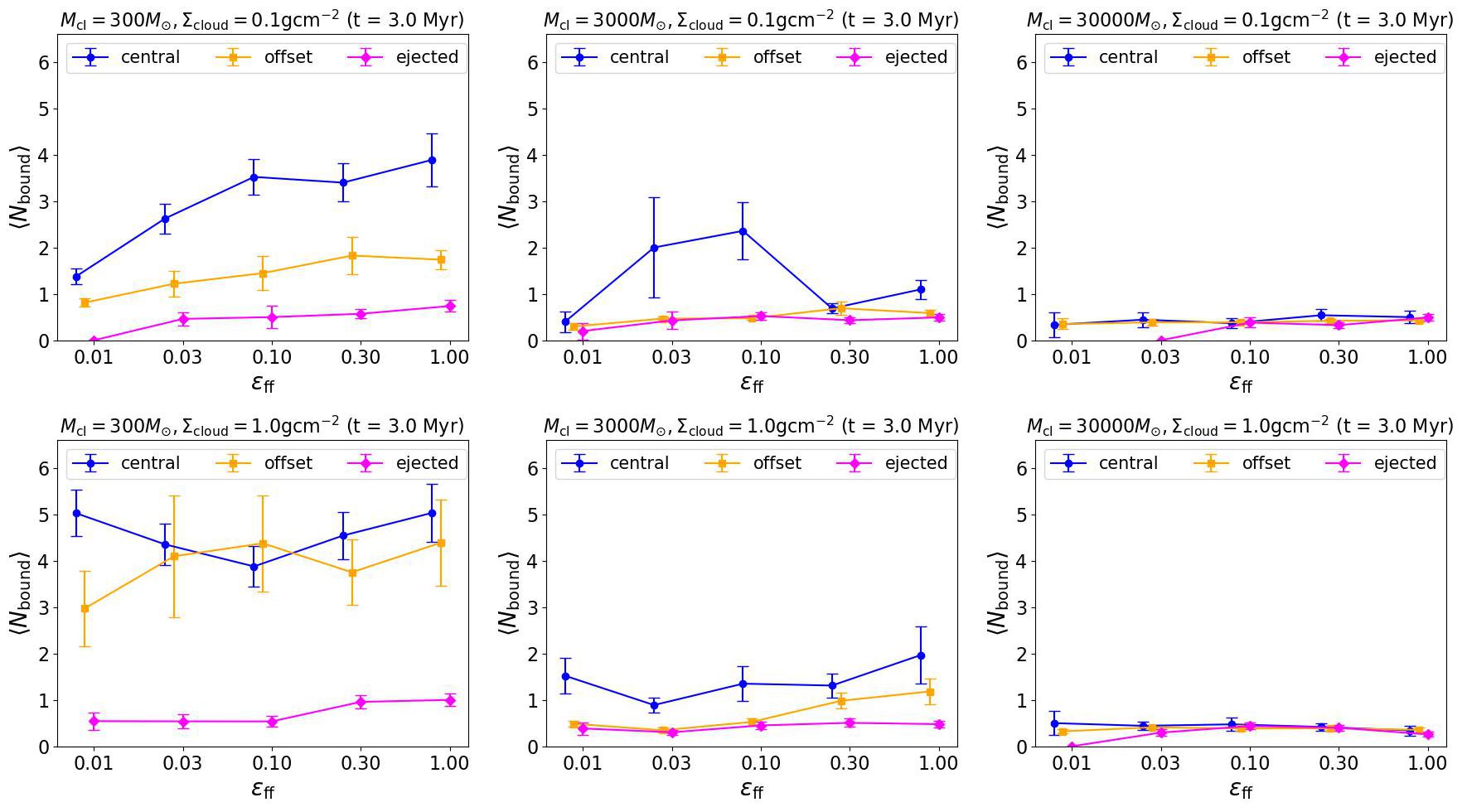}
    \caption{Each panel shows comparison between different \sfeff models showing \avgNcomp around massive stars at cluster age of $t = 3.0 $ Myr. Panels in the first and second rows show low ($\Sigmacloud = 0.1\: \rm g\: cm^{-2}$) and high ($\Sigmacloud = 1.0\: \rm g\: cm^{-2}$) cloud mass surface density cases, respectively. From left to right, the columns show M300, M3000, and M30000 models, respectively. Each panel shows \avgNcomp around massive stars in three subsets: \textit{central} stars (blue), \textit{offset} stars (orange) and \textit{ejected} stars (magenta). $\langle ... \rangle$ indicates averaging over massive stars in all realizations of each model. The points for different subsets are offset by [0.1,0,-0.1] along $x-$axis to allow for visual comparison. We find that massive stars in different \sfeff clusters have variations in number of bound companions at the same cluster age, possibly due to variations in the star cluster's dynamical history.}
    \label{fig:model_comparison_sfeff_avgNbound}
\end{figure*}

\begin{figure*}
    \centering
    \includegraphics[width=0.95\textwidth]{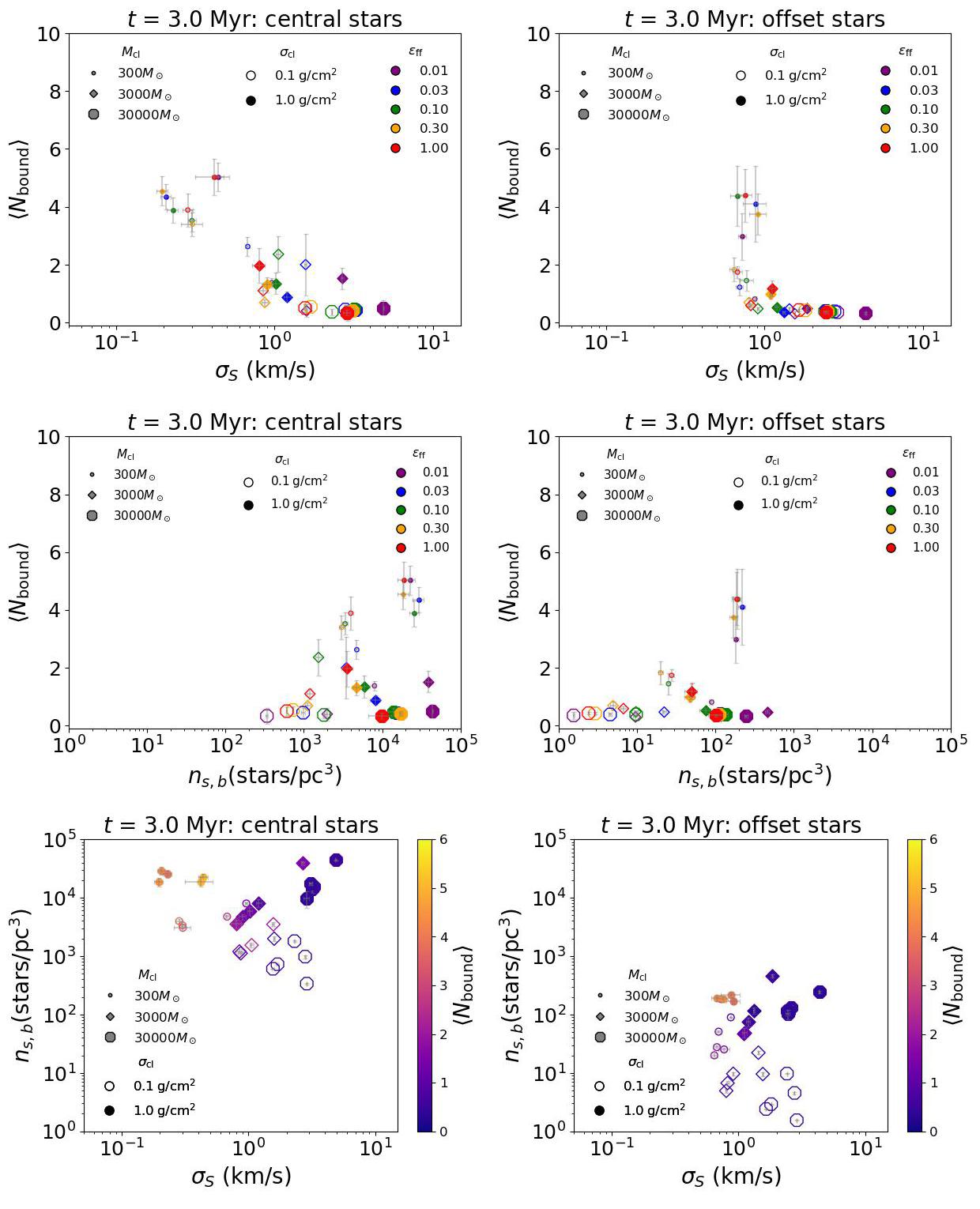}
    \caption{Comparison between different (\Mcl,\Sigmacloud,\sfeff) models showing how \avgNcomp around \textit{central} stars (left column) and \textit{offset} stars (right column) is affected by the stellar velocity dispersion (\veldisstar) and stellar number density (\ndensstar) within their respective radial regions in the bound star cluster. The first and second rows show \avgNcomp for each model versus the corresponding \veldisstar and \ndensstar, respectively, at a cluster age of $t$ = 3.0 Myr. The gray lines show the standard errors on the means calculated from the multiple realizations ($N_{\rm sims}$) of each model. In the third row, a scatter plot of \veldisstar and \ndensstar for each (\Mcl, \Sigmacloud) model is color-coded by the value of \avgNcomp for that model. The colorbar shows the range of \avgNcomp. We find that in star clusters with lower velocity dispersion and higher stellar densities, massive stars will gather more bound companions.}
    \label{fig:environment_effect_on_Nbound}
\end{figure*}

Figure~\ref{fig:model_comparison_sfeff_avgNbound} shows \avgNcomp around massive stars at cluster age of $t = 3.0 $ Myr compared across \sfeff models. We find that \textit{central stars} have larger \Ncomp than \textit{offset} and \textit{ejected} stars across almost all models. We also notice there is a gradual rise in \Ncomp with \sfeff in the M300L model. However, the variation of \Ncomp with \sfeff is not monotonous in other models. Nonetheless, this hints that the number of bound companions around a massive star can be affected by the formation timescales of its star cluster. 

This is primarily due to the time evolution of stellar velocity dispersion (\veldisstar) and stellar number density (\ndensstar) around the massive star during the formation of the star cluster. Figure~\ref{fig:environment_effect_on_Nbound} shows the variation of \avgNcomp with these environmental variables for both \textit{central} stars (left column) and \textit{offset} stars (right column). For \textit{central} stars, we calculate \veldisstar and \ndensstar in the central core, within 20\% of the half-mass radius, of the bound cluster. For \textit{offset} stars, we calculate \veldisstar and \ndensstar in a spherical shell extending from 0.2 to 3.0 times the half-mass radius. We observe that the stellar densities (\ndensstar) are smaller around \textit{offset} stars than \textit{central} stars, as expected by their location. 

For both \textit{central} and \textit{offset} stars, we find that \avgNcomp is greater for smaller values of stellar velocity dispersion (\veldisstar, see first row). The dependence of \avgNcomp on stellar density (\ndensstar) is more complex (see second row). If we check the third row, we find that \avgNcomp is maximum for models with lower velocity dispersion and higher stellar density (tending towards the upper left corner). These happen to be the M300H models with the highest \avgNcomp, as also seen in Figure~\ref{fig:model_comparison_variation_Mprim_Mcomp_fullset}. This is true for both \textit{central} and \textit{offset} stars. This shows that the prevailing stellar velocity dispersion and number density around a massive star in a forming star cluster can determine the number of bound companions it gathers. 

The reasoning is as follows: bound systems of multiple stars form and disrupt through stellar encounters. The size of these bound systems will be determined by the balance of captures and disruptive encounters. The number of possible interactions will depend on the number density (\ndensstar) around the massive star, because it determines the availability of stars which could undergo such encounters (both captures and disruptions). However, the velocity dispersion (\veldisstar) of the environment determines whether captures or disruptions win. This is because \veldisstar controls the strength of the gravitational focusing ($v_{\rm max}/\veldisstar$) experienced by companions during capture. A smaller \veldisstar environment increases the gravitational focusing by increasing the impact parameter. In denser environments with larger velocity dispersions, captures become more difficult, as encounters are on average more energetic. Among our models, the low \Mcl clusters provide environments that favor captures over disruptions because of their lower velocity dispersions, and thus M300 models have higher \avgNcomp.

We note that dynamical mass segregation does contribute to the increase in \avgNcomp, as more massive stars move to the more crowded center (higher \ndensstar) with more potential companions that could be captured. While we have found mass segregation to be more efficient in our low \Mcl clusters (see Paper II, Section 4.3.1), \avgNcomp will depend more on the \veldisstar of the environment rather than mass segregation alone. This is because velocity dispersion ultimately determines the balance of captures that form the multiple systems versus the encounters that disrupt them.

\begin{figure*}
    \centering
    \includegraphics[width=\textwidth]{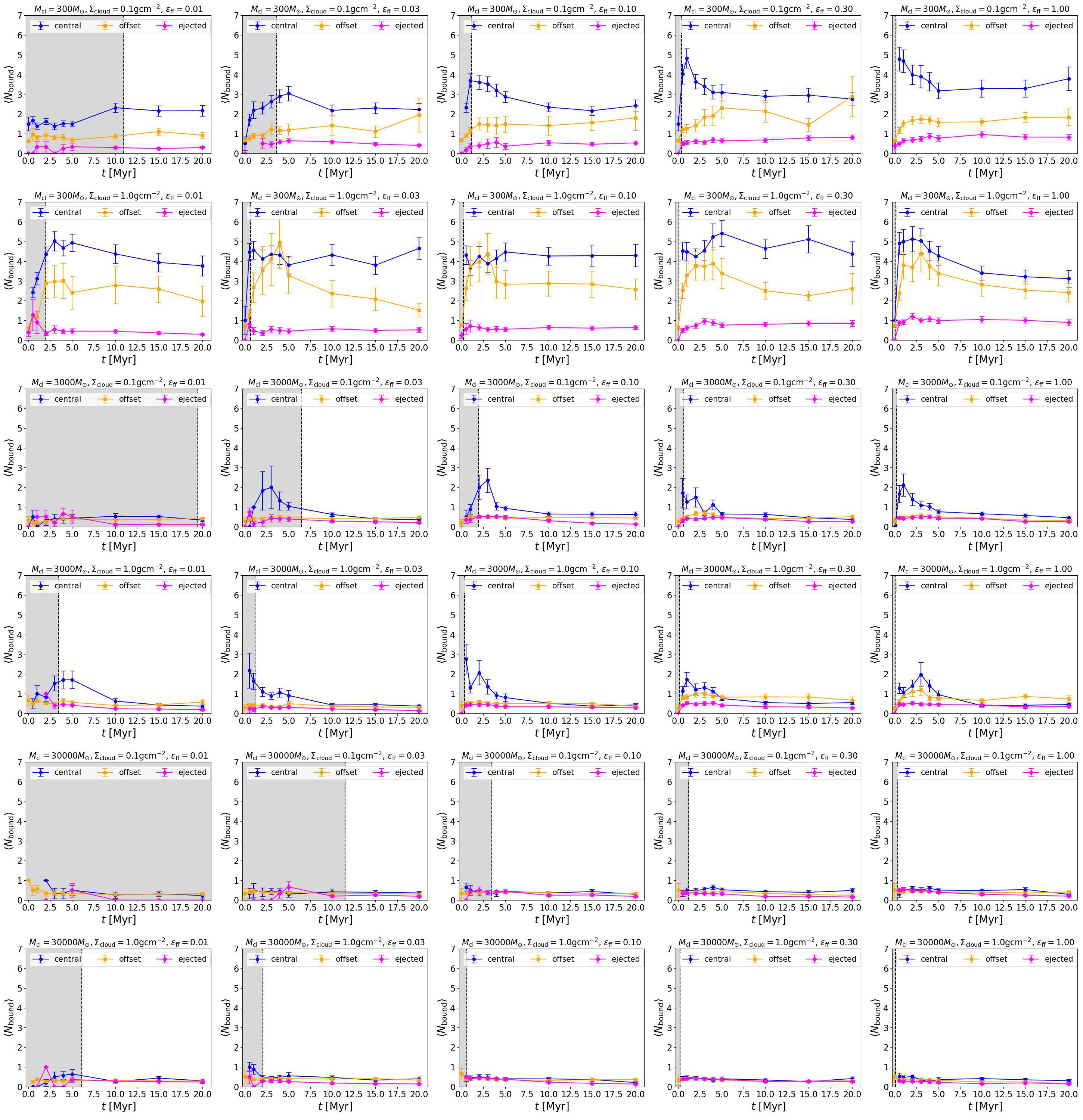}
    \caption{Comparison between different models (\Mcl,\Sigmacloud,\sfeff) across our full set showing the evolution of average number of bound companions \avgNcomp around massive stars versus cluster age $t$ in Myr. The columns show the five different values of \sfeff. The rows (from top to bottom) show M300L, M300H, M3000L, M3000H, M30000L, and M30000H sets, respectively. Each panel shows \avgNcomp around massive stars in three subsets: \textit{central} stars (blue), \textit{offset} stars (orange) and \textit{ejected} stars (magenta). $\langle ... \rangle$ indicates averaging over massive stars in all realizations of each model. The dashed black vertical line represents the formation timescale (\tform) of the model. The star cluster is still embedded in background gas before \tform, represented by the shaded region. We find that the number of bound companions around massive stars will be affected by the stellar environment (\veldisstar and \ndensstar) of their star cluster during its formation, gas expulsion and subsequent expansion.}
    \label{fig:model_comparison_time_evolution_avgNbound_all}
\end{figure*}

Variations in these environmental variables (\veldisstar and \ndensstar) in the same cluster and between star clusters forming at different timescales leads to the variation in \avgNcomp between massive stars at a given cluster age. This also means that as the cluster forms and evolves, \avgNcomp will evolve with time due to changes in \veldisstar and \ndensstar. In Figures~\ref{fig:model_comparison_time_evolution_avgNbound_all}, we show the evolution of \avgNcomp around massive stars with cluster age $t$ in Myr across our full set of models (\Mcl,\Sigmacloud,\sfeff). Each model has its own formation timescale (\tform) indicated by the dashed vertical line in each panel. The star cluster is still embedded in background gas before \tform. We find that across all models, there is a gradual rise in \Ncomp before a decline and/or flattening as the cluster ages. This indicates that the formation timescale and stellar environment (density and velocity dispersion) have a significant effect on \avgNcomp.  

\begin{figure*}
    \centering
    \includegraphics[width=\textwidth]{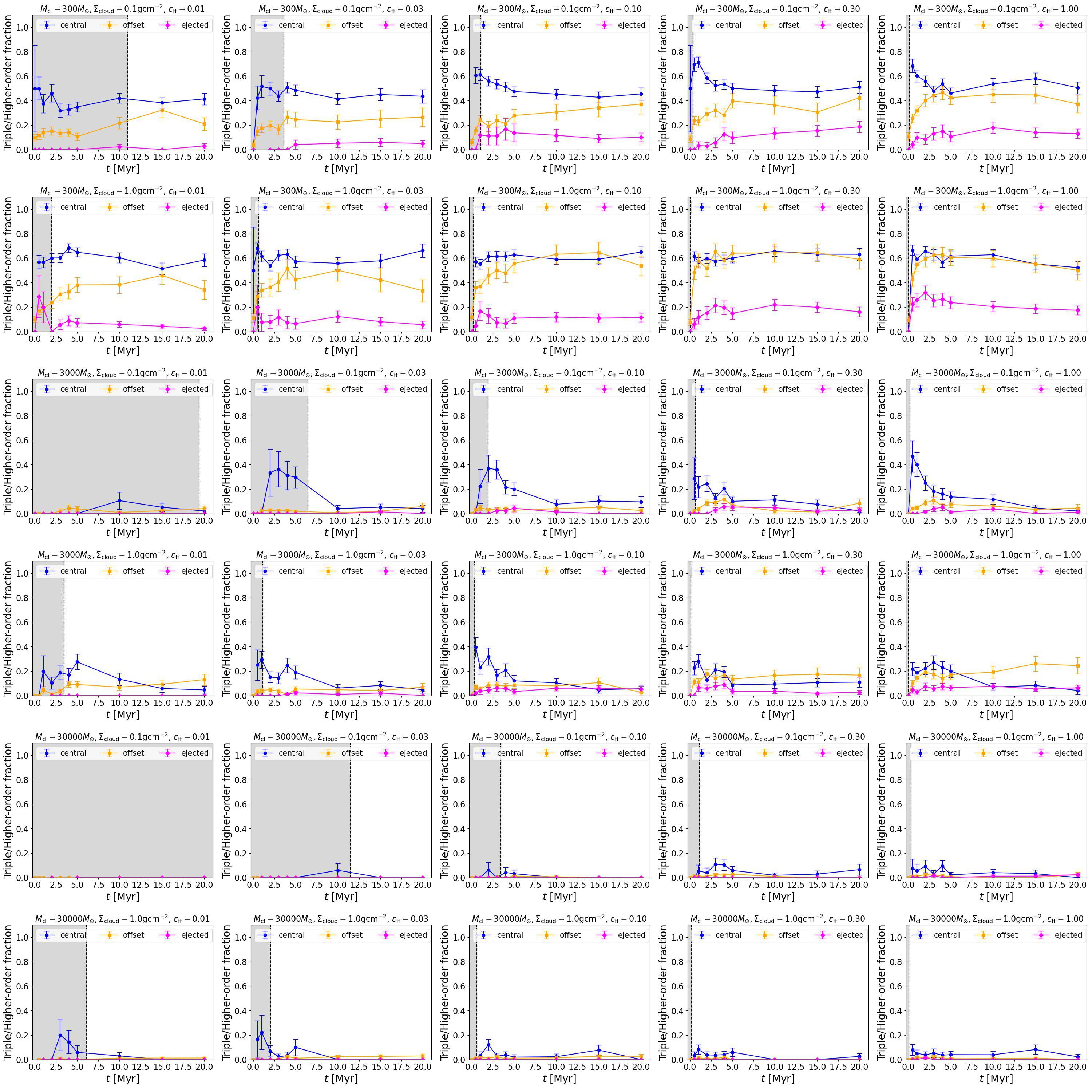}
    \caption{Comparison between different models (\Mcl,\Sigmacloud,\sfeff) across our full set showing the evolution of Triple/Higher-order fraction (THF) of massive stars versus cluster age $t$ in Myr. The columns show the five different values of \sfeff. The rows (from top to bottom) show M300L, M300H, M3000L, M3000H, M30000L, and M30000H sets, respectively. Each panel shows THF of massive stars in three subsets: \textit{central} stars (blue), \textit{offset} stars (orange) and \textit{ejected} stars (magenta). THF was calculated by aggregating massive stars across all realizations of each model and by counting the fraction of massive stars with triple or higher-order bound companions. The dashed black vertical line represents the formation timescale (\tform) of the model. The star cluster is still embedded in background gas before \tform, represented by the shaded region. We find that bound systems with three or more stars can form dynamically in a forming star cluster. The rise and decline of the fraction of these systems depends on the evolving stellar environment (\veldisstar and \ndensstar) of the star cluster.}
    \label{fig:model_comparison_time_evolution_thf_all}
\end{figure*}

Besides \avgNcomp, the build-up of bound stellar systems around massive stars is also reflected in the triple/higher-order fraction of massive stars (THF, Equation \ref{eqn:thf}). Our simulations have only 50\% primordial binaries, and triple/higher-order companions are gathered dynamically over time. These companions can be lost or replenished over time in the star cluster environment, so THF evolves over time. Figure~\ref{fig:model_comparison_time_evolution_thf_all} shows the time evolution of THF for the full set of our models (\Mcl, \Sigmacloud, \sfeff). We find that for \textit{central} and \textit{offset} stars, THF usually rises before flattening and/or declining as the star cluster ages. Depending on the time evolution of \veldisstar and \ndensstar, and thus dynamical captures of the model, this rise in THF occurs quickly in the first Myr whereas for others, it takes more time. There is also variation across models in the flattening or fall of THF once gas expulsion is complete. M300 models have higher THF which reflects their larger \avgNcomp. In these smaller \Mcl clusters, THF rises quickly and flattens. In contrast, THF in M3000 clusters rises and then falls gradually over time. In the M30000 clusters, triple and higher-order bound systems are rarely formed, so THF remains at a small level and does not change much over time (THF < 0.2). Finally, we note that the evolution of THF around \textit{offset} stars is more gradual with the THF levels mostly flattening over time, whereas for \textit{central} stars, there can be flattening or decline. THF remains low around \textit{ejected} stars across all models. 

\subsection{Evolution of projected stellar density profiles}
\label{subsec:results-projectedstellardensities}

\begin{figure*}
    \centering
    \includegraphics[width=0.76\linewidth]{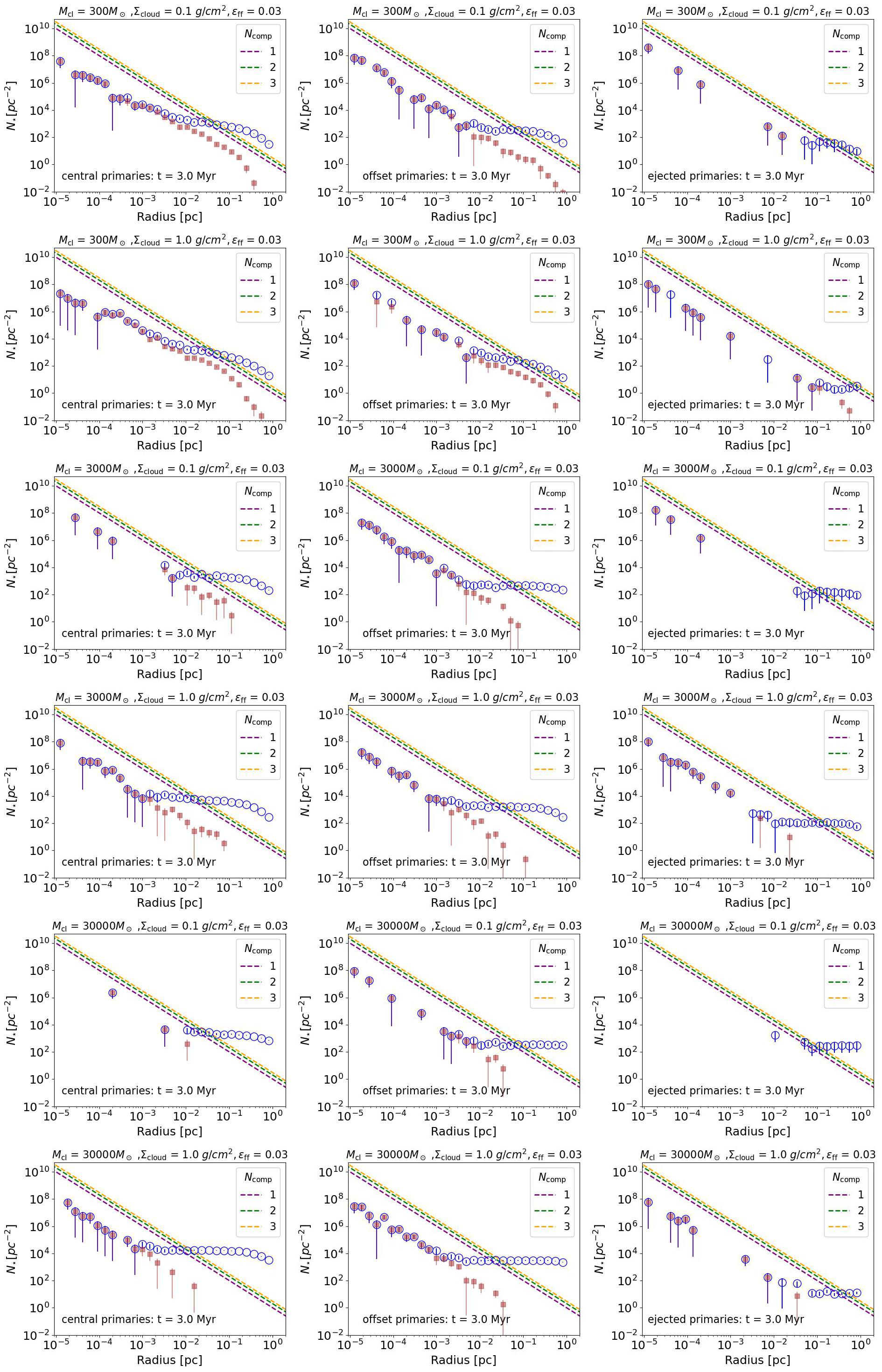}
    \caption{Each panel shows averaged projected stellar density profiles ($\langle N_\star (r) \rangle$) around massive stars in the fiducial (\sfeff = 0.03) model at cluster age of $t$ = 3.0 Myr. The rows show different sets in the following order from top to bottom: M300L, M300H, M3000L, M3000H, M30000L and M30000H, respectively. The columns (from left to right) show the profiles around \textit{central} stars, \textit{offset} stars, and \textit{ejected} stars, respectively. In each panel, the brown points include only the bound companions, whereas the blue points include all nearby companions. The errorbars are the standard errors on the $\langle N_\star (r) \rangle$ in the bin at each radius $r$. The dashed lines show the expected projected stellar densities if the bin had 1, 2, or 3 stars, respectively. The stars are counted in logarithmically spaced annular bins ($N_{\rm bins}$ = 30). We find that the projected stellar density profile of all companions is relatively flat, while the bound companions are found close ($r$ < 0.1 pc) to the massive star.}
    \label{fig:massivestar_nstar_profile}
\end{figure*}

In Figure~\ref{fig:massivestar_nstar_profile}, we show the averaged projected stellar density profiles around \textit{central}, \textit{offset} and \textit{ejected} massive stars. We first constructed projected stellar density profiles ($N_\star (r)$) around every primary massive star as described in section~\ref{subsec:methods-projectedstellardensities}. Then, we averaged the individual projected stellar density profiles over all massive stars in each subset to generate an averaged projected stellar density profile shown in the figure. We find that in all fiducial models, there is a background density of stars, the blue profile, which includes all stars surrounding the massive star. The blue profile remains mostly flat at large $r$ however, near the massive star, there is a gradual rise in $N_\star$ in the innermost bins ($r < 0.01 \rm pc$). This is due to the presence of bound companions near the massive star, indicated by the brown profile. However, we find that the brown profile quickly drops off beyond 0.1 pc, contributing only a very small fraction of the total $N_\star$. We note that the bound profile stretches out to further distances around massive stars in M300 models. This is consistent with the fact that these models have small velocity dispersion and thus massive stars can gather more companions, possibly at larger separations, leading to more extended bound stellar systems. Finally, we also note that the background projected stellar density (blue profile) around \textit{ejected} stars is lower in comparison to the \textit{central} and \textit{offset} stars, as expected from their location in the cluster. The bound companions (if any) around these \textit{ejected} massive stars show up as the rise in $N_\star$ in the innermost bin. 

\begin{figure*}
    \includegraphics[width=\textwidth]{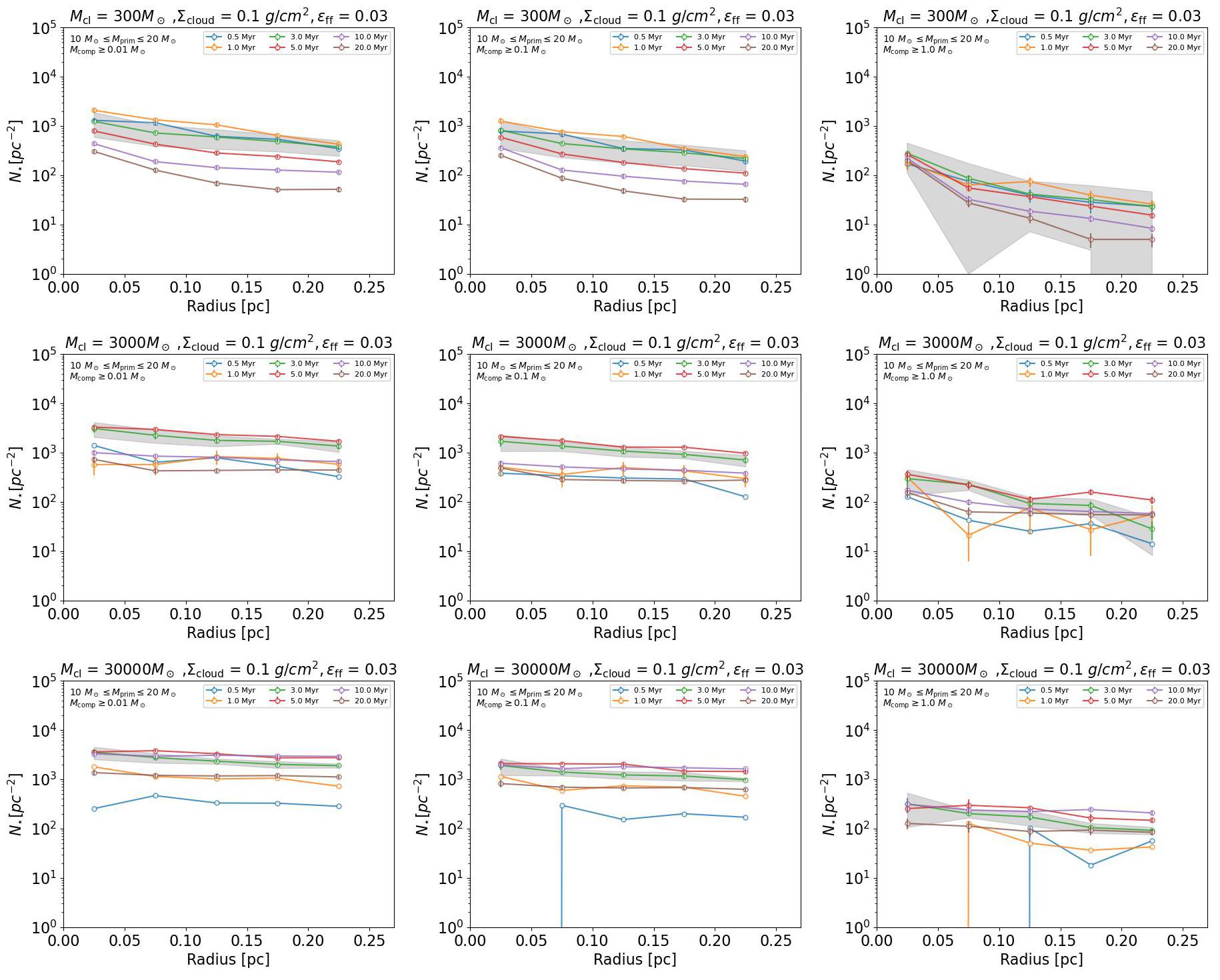}
    \caption{Each panel shows the time evolution of projected stellar density profiles ($\langle N_\star (r) \rangle$) around massive stars in simulated clusters for the fiducial (\sfeff = 0.03) case in low cloud mass surface density ($\Sigmacloud = 0.1\: \rm g\: cm^{-2}$) sets. The rows show M300L, M3000L and M30000L models, respectively. The columns show the results considering companions with masses above $0.01 \msun$, $0.1 \msun$ and $1.0 \msun$, respectively. In every panel, each line represents averaged values of projected stellar density around \textit{central} stars in the 10 \msun to 20 \msun mass range at the given cluster age, across multiple realizations for that model. The errorbars represent the standard errors. The shaded region represents the standard deviation for the $t$ = 3.0 Myr profile. We find that the projected stellar density profile around a massive star rises during ongoing star formation, reaches a maximum before gas expulsion, and declines as the star cluster expands following gas removal.}
    \label{fig:projectedstellardensities_lowsigma}
\end{figure*}

\begin{figure*}
    \includegraphics[width=\textwidth]{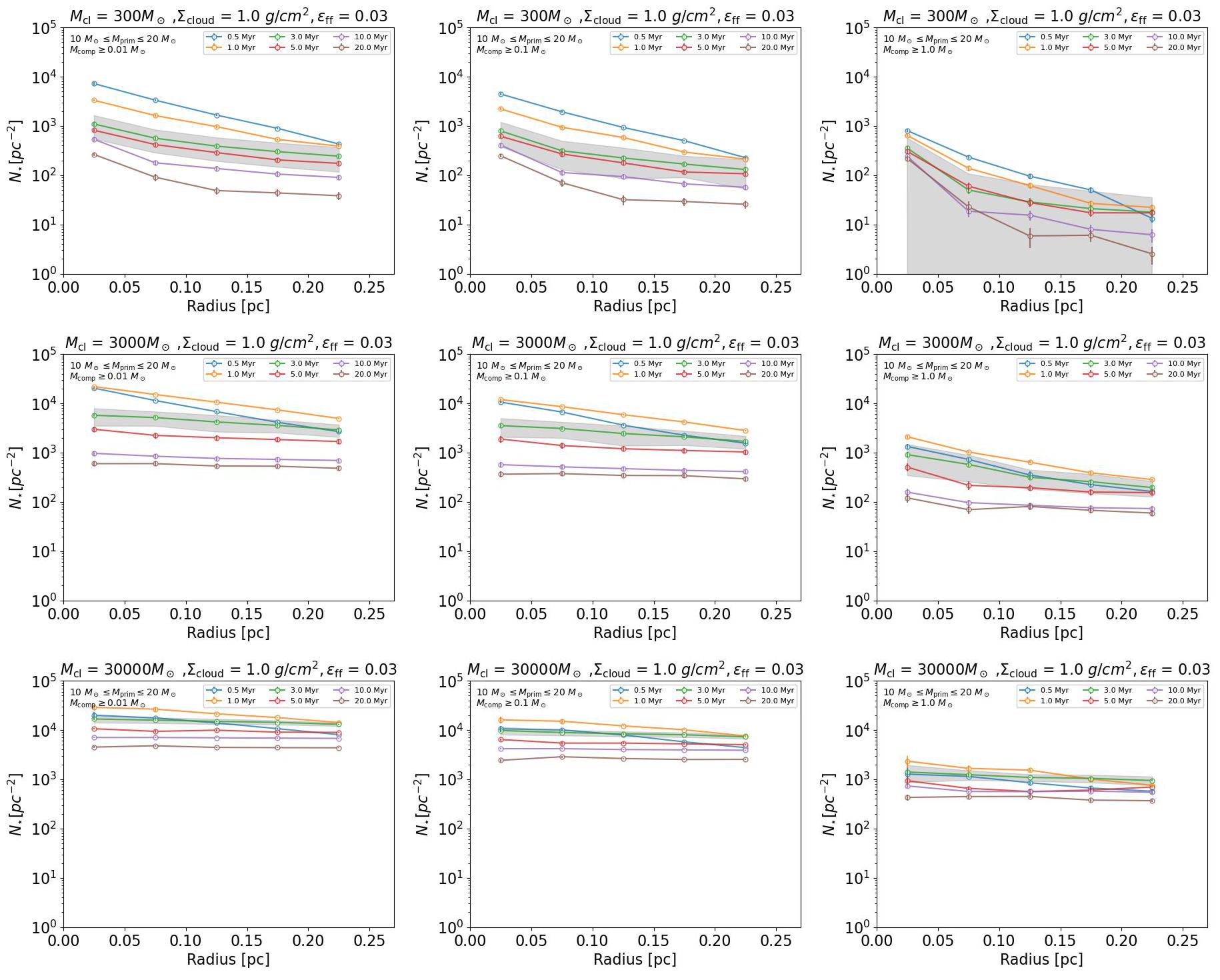}
    \caption{Each panel shows the time evolution of projected stellar density profiles ($\langle N_\star (r) \rangle$) around massive stars in simulated clusters for the fiducial (\sfeff = 0.03) case in high cloud mass surface density ($\Sigmacloud = 1.0\: \rm g\: cm^{-2}$) sets. The rows show M300H, M3000H and M30000H models, respectively. The columns show the results considering companions with masses above $0.01 \msun$, $0.1 \msun$ and $1.0 \msun$, respectively. In every panel, each line represents averaged values of projected stellar density around \textit{central} stars in the 10 \msun to 20 \msun mass range at the given cluster age, across multiple realizations for that model. The errorbars represent the standard errors. The shaded region represents the standard deviation for the $t$ = 3.0 Myr profile. In addition to the time evolution, observational limits also affect the shape of the projected stellar density profiles.}
    \label{fig:projectedstellardensities_highsigma}
\end{figure*}

To compare our projected stellar density profiles with observations in section~\ref{subsubsec:results-observations-nstar}, we considered \textit{central} massive stars in the mass range from 10 to 20 \msun. Then, we constructed projected stellar density profiles around these stars at different cluster ages to study their time evolution as the cluster is formed. Figures~\ref{fig:projectedstellardensities_lowsigma} and \ref{fig:projectedstellardensities_highsigma} show time evolution of \avgNstar profiles for different \Mcl models (rows) in low and high \Sigmacloud cases, respectively. These profiles include all bound and unbound companions around the massive stars. To reproduce the effect of observational detection limits, we vary the companion mass threshold between $\minmasscomp \geq 0.01 \msun$, $\minmasscomp \geq 0.1 \msun$ and $\minmasscomp \geq 1.0 \msun$ across the columns. In the rows, we compare different \Mcl models for the fiducial \sfeff = 0.03 case.

Let us consider the first column in each figure. We find that across each model, the \avgNstar profile around $t$ = 0.5 Myr starts at a baseline level determined by the mass (hence, number of stars formed) and size of the star cluster, which are set by the initial conditions (\Mcl,\Sigmacloud,\sfeff). For instance, the M30000 models are more compact and have a higher initial \avgNstar than M300 or M3000 models. In general, the profile rises with time as more stars are formed and the massive stars, which have already formed, gather companions around them. This continues as long as the cluster is embedded in its gas clump, which is gradually dissipated away. Once the gas is depleted at t = \tform, the \avgNstar profile starts to decline. This is because of the loss of binding potential provided by the gas clump, which causes the star cluster to expand and the projected stellar density around each massive star to fall. The time evolution of the \avgNstar profile is thus affected by the formation timescale of the star cluster. 

Now, if we consider the same \Mcl model (row) and compare the columns, we find that on increasing the minimum companion mass, the \avgNstar declines because we miss the lower-mass companions around the massive star. The fall in \avgNstar is around a factor of 10 between $\minmasscomp \geq 0.1 \msun$ and $\minmasscomp \geq 1.0 \msun$, and a factor of 2 between $\minmasscomp \geq 0.01 \msun$ and $\minmasscomp \geq 0.1 \msun$. This indicates that observational studies of stellar clustering around massive stars should be able to detect low-mass companions as small as 0.01 \msun to properly test models of high-mass star formation. Conversely, any comparison of simulated models with observations should consider both the cluster age and detection limit, considering that \avgNstar is affected by time evolution and the detected companion masses, respectively.

\subsection{Comparison with observations}
\label{subsec:results-observations}

\subsubsection{Comparison of high-mass star formation models}
\label{subsubsec:results-observations-nstar}

We compare the projected stellar density profiles (\avgNstar) in our simulations with observations in this section. We consider $N_\star (r)$ observations around high-mass stars in two star forming regions: a massive protostar in AFGL 5180, studied by \cite{Crowe2024A&A...682A...2C}, and a high-mass star-forming complex G19.88-0.53, studied by \cite{CostaSilva2022A&A...659A..23C}. For these previous observations, the surface number density was derived by counting the number of YSOs within a given annulus, each annulus of width 0.05 pc. For a fair comparison of simulations with observations, we have counted companions around massive stars in the same linearly spaced bins. Since we are studying the projected number density profiles of stars, we choose the final stellar mass (\finalstellarmass) as the common variable when comparing the variation of \avgNstar between simulations and observations during our analysis.

The total gas mass of the AFGL 5180 star-forming clump is estimated from spectral energy distribution (SED) fitting to be around 300 \msun to 500 \msun in the literature \citep{Maity2023MNRAS.523.5388M, Telkamp2025ApJ...986...15T}. For our analysis, we assume the gas mass of AFGL 5180 clump to be $M_{\rm cl}$ = 400 \msun. Since we do not know the actual final star-formation efficiency ($\epsilon$) of the AFGL 5180 region, we assume a range of $\epsilon$ = [0.1, 0.3, 0.5, 0.8] to estimate the final stellar mass ($M_{\rm stars} = \epsilon M_{\rm cl}$) expected to form in the region. The rough lower limit of \finalstellarmass = 40 \msun ($\epsilon = 0.1$) is set because \citet{Telkamp2025ApJ...986...15T} identified four protostars in AFGL 5180 by SED fitting: three in the range 12-16 \msun and one with 5 \msun. The rough upper limit of \finalstellarmass = 320 \msun ($\epsilon = 0.8$) assumes that 20\% of gas mass is lost due to effects of stellar feedback. So, the order of magnitude estimate of \finalstellarmass for AFGL~5180 is around 100 \msun. Among the four protostars, the best-fit mass of central massive protostar (designated S4) in AFGL~5180 is inferred to be $11.3^{+6.1}_{-4.0} \msun$ by SED fitting \citep{Crowe2024A&A...682A...2C}. The observed \avgNstar profile was calculated by those authors around this protostar. So, we constructed averaged projected stellar density profiles around massive stars in the 10 \msun to 20 \msun mass range across multiple realizations. These observational studies had a detection limit of $0.1 \msun$. So, we consider profiles generated for the case with $\minmasscomp \geq 0.1 \msun$. However, for completeness, we also include the cases of $\minmasscomp \geq 0.01 \msun$ and $\minmasscomp \geq 1.0 \msun$ in Table~\ref{tab:powerlawfits}.

On the other hand, the G19.88-0.53 region is likely a proto-stellar cluster where multiple massive stars are forming and driving the numerous bipolar outflows observed in the radio and near-infrared \citep{Issac2020MNRAS.497.5454I, CostaSilva2022A&A...659A..23C}. Nonetheless, by assuming a single protostellar source dominates the luminosity of the region, the authors performed SED fitting to derive best-fit masses ranging from 4 to 8 \msun for a proto-massive star forming in an initial gas core of mass ranging from 100 to 480 \msun, respectively. They also derived a YSO number density of $170 \pm 46.5 \rm \: pc^{-2}$ in the region, which we include in our comparison. For completeness, we note that the above SED fitting is performed with radiative transfer models from \citet{ZhangTan2018ApJ...853...18Z}, which are based on the Turbulent Core Accretion (TCA) model for massive star formation \citep{McKee2003ApJ...585..850M}. The formation of high-mass stars in our TCCA simulations can be considered to be occurring stochastically via TCA within our clumps.

\begin{figure*}
    \centering
    \includegraphics[width=0.9\linewidth]{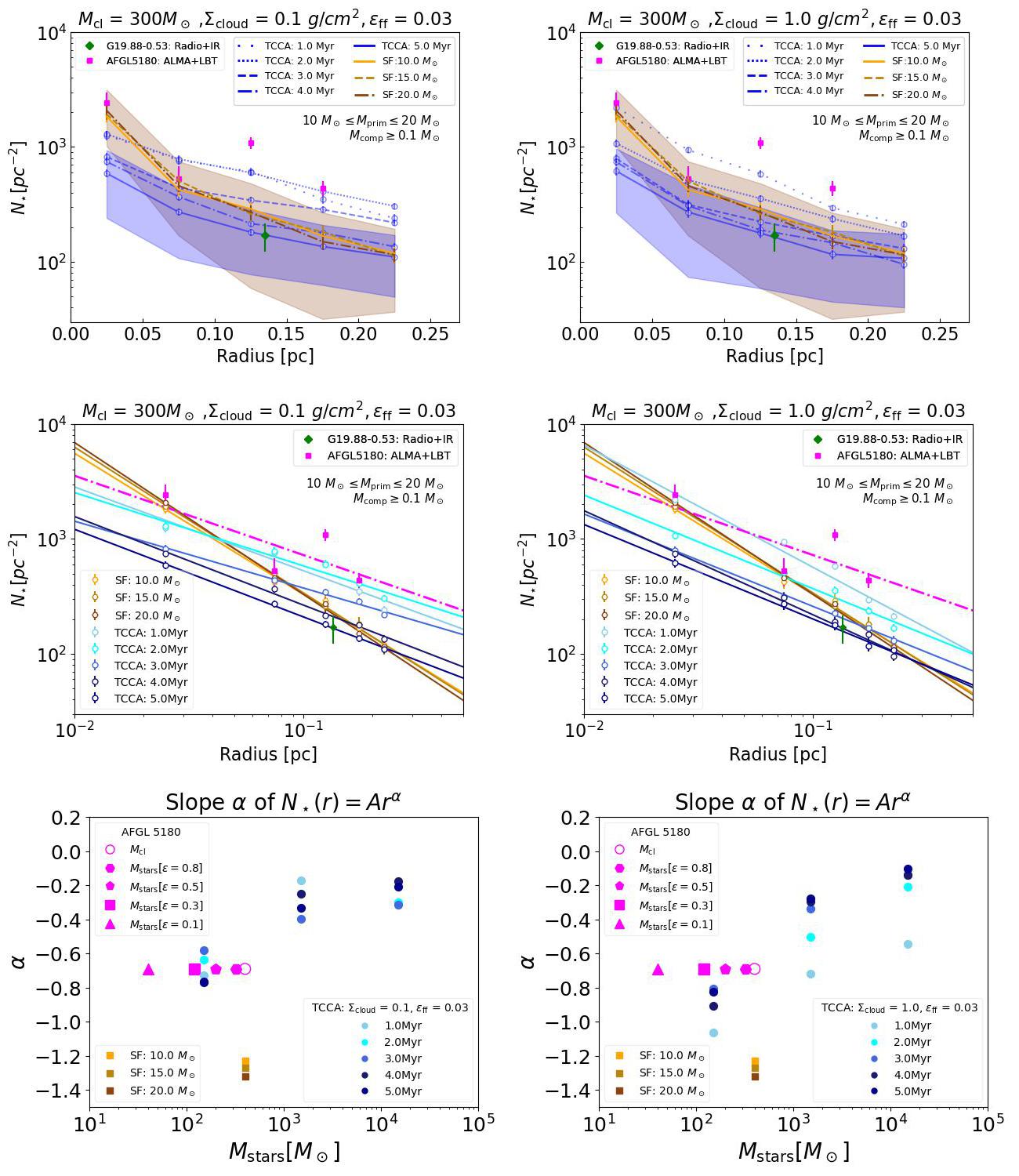}
    \caption{\textbf{Top row}: Projected stellar density profiles for our fiducial (\sfeff = 0.03) M300L and M300H models (TCCA models, shown in blue), which were found to be the best match to the observations from AFGL 5180 (magenta squares) and G19.88-0.53 (green diamond). Blue lines with different linestyles represent TCCA profiles at different cluster ages. The errorbars represent the standard errors. The blue shaded region represents the standard deviation for the $t$ = 5.0 Myr profile. These profiles were created around \textit{central} massive stars which fall in the mass range $10-20 \msun$ and their companions have masses $\minmasscomp \geq 0.1 \msun$, matching the observational limits. We also compare with simulated profiles averaged around massive stars from \STARFORGE simulations in Grudic et al. 2022. Orange lines of different shades represent the profiles when the massive star reaches different masses. The orange shaded region represents the standard deviation for the case of 20 \msun profile. \textbf{Middle row}: Power-law fits of the form $\avgNstar (r) = A r^{\alpha}$ to the projected stellar density profiles presented in the top row. Blue lines of different shades represent fits for our TCCA models at different cluster ages. Orange lines of different shades represent fits for \STARFORGE models when the massive star reaches different masses. The magenta line represents the fit to AFGL 5180 observations. The constants $A$ and $\alpha$ for these power-law fits are tabulated in Table~\ref{tab:powerlawfits}. \textbf{Bottom row}: Index ($\alpha$) of the power-law fits presented in the middle row versus the final stellar mass (\finalstellarmass) formed in each model. TCCA models form 0.5\Mcl in stellar mass and this particular \STARFORGE model also forms 400 \msun of stars. The AFGL 5180 clump has a mass of $M_{\rm cl}$ = 400 \msun: we estimate \finalstellarmass for AFGL 5180 assuming different $\epsilon$, where $\finalstellarmass = \epsilon M_{\rm cl}$. We find that the observed AFGL 5180 projected stellar density profile has a flat shape consistent with our TCCA models.}
    \label{fig:bestfitmodelNstarprofile}
\end{figure*}

In the top row of Figure~\ref{fig:bestfitmodelNstarprofile}, we present projected stellar density profiles in our M300L and M300H models, which we found to be the best-fit models to the observations, across the parameter space of our models (\Mcl, \Sigmacloud, \sfeff) and cluster age ($t$). Our TCCA simulations form a star cluster with \finalstellarmass = 0.5 \Mcl, so M300 models have \finalstellarmass = 150 \msun. Since we are comparing with young star-forming regions, we examined snapshots at $t$ = 1.0, 2.0, 3.0, 4.0 and 5.0 Myr in our simulations. As a result of our multiple realizations for each model, we are able to obtain a reasonably large set of massive stars in the 10-20 \msun mass range so that we can create averaged profiles at each cluster age.

For comparison to other types of models, we include projected stellar density profiles constructed around massive stars in the \STARFORGE simulations from \cite{Grudic2022MNRAS.512..216G}. Their simulations model the collapse of a giant molecular cloud with an initial mass of 20,000 \msun and size of 10 pc into a young stellar cluster with the formation of protostars as sink particles and incorporating their feedback when they become stars. At $\epsilon$ = 0.02, this \STARFORGE simulation forms a cluster with \finalstellarmass = 400 \msun \citep{Grudic2022MNRAS.512..216G,Crowe2024A&A...682A...2C}, which has the closest resemblance to both AFGL 5180 ($\sim$ 100 \msun) and M300 models of TCCA (150 \msun) among the publicly available \STARFORGE simulation runs. As mentioned above, we assume \finalstellarmass as the common variable when comparing models and observations. The formation of massive stars in \STARFORGE simulations resemble the Competitive Accretion model of high-mass star formation. In the case of \STARFORGE, massive stars build up their mass over time by gas accretion while competing against neighboring low-mass companions. So, \STARFORGE simulations had massive stars in the required mass range of 10 \msun to 20 $\msun$ only from $t$ = 3.0 Myr onwards. At early cluster ages (3.0 Myr $\rm \leq t \leq $ 6.0 Myr), only one or two protostars reach this mass threshold. If we generated profiles at these ages, they would be highly subject to random scatter because we do not have a significant number of massive stars in the required mass range. This is further complicated by the fact that the \STARFORGE model has only one realization, so we can only analyze the set of 28 protostars that reach this mass range at some point during this simulation. Therefore, we chose to analyze projected number density profiles when these massive stars reached a certain mass instead. Then, we average the profiles around the full set of 28 massive stars that reach the same 10 \msun to 20 \msun mass range during their formation. In Figure~\ref{fig:bestfitmodelNstarprofile}, we present the averaged projected stellar density profiles expected when a massive star reaches masses of 10 \msun, 15 \msun and 20 \msun in \STARFORGE simulations. 

To quantify the shape of the \avgNstar profiles, we fit power-laws of the form: $\avgNstar (r) = A r^{\alpha}$. The calculated values of $A$ and $\alpha$ for each profile are presented in Table~\ref{tab:powerlawfits}. The columns for the $\minmasscomp \geq 0.1 \msun$ case represent the profiles presented in the middle row of Figure~\ref{fig:bestfitmodelNstarprofile}. However, we also explore the effect of observational limits on \minmasscomp on the shape of \avgNstar profiles for the other two cases ($\minmasscomp \geq 0.01 \msun$ and $\minmasscomp \geq 1.0 \msun$), and present their fits in Table~\ref{tab:powerlawfits}. While the first two rows of Figure~\ref{fig:bestfitmodelNstarprofile} focus on M300 models, we also present fits for M3000 and M30000 models in Table~\ref{tab:powerlawfits}.   

From Figure~\ref{fig:bestfitmodelNstarprofile}, we find that there is a wide scatter in both our and \STARFORGE projected stellar density profiles (indicated by the shaded regions). However, profiles from \STARFORGE models are generally steeper than our profiles, and fall quickly beyond $r > 0.1 \rm\: pc$. In contrast, our projected stellar density profiles remain relatively flat out to 0.25 pc, which is also seen in Figures~\ref{fig:projectedstellardensities_lowsigma} and \ref{fig:projectedstellardensities_highsigma}. The \STARFORGE profiles for the 10 \msun, 15 \msun and 20 \msun cases are quite similar even though they get slightly steeper as the massive star accretes mass to rise from 10 \msun to 20 \msun (see power-law indices ($\alpha$) in columns 5-6 of Table~\ref{tab:powerlawfits}). We also note the time evolution of our models in Figure~\ref{fig:bestfitmodelNstarprofile}. The profile rises in the first Myr as new stars form in the vicinity of the massive star. The star formation is complete by $\tform$ = 3.64 Myr and 0.65 Myr in the M300L and M300H models, respectively (see Table~\ref{tab:simulations}). Then, the star cluster expands following gas removal. Thus, \avgNstar profile decreases with time. If we consider all low-mass companions ($\minmasscomp > 0.01 \msun$), we find that our profiles at $t$ = 5.0 Myr are in general flatter than corresponding early profiles at $t$ = 1.0 Myr. We also note the effect of observational limits (\minmasscomp) on the shape of profiles. We find that on increasing \minmasscomp from 0.01 \msun to 1.0 \msun, the profile steepens for \STARFORGE profiles. This is generally true for our models as well (refer power-law indices ($\alpha$) in Table~\ref{tab:powerlawfits}).   

Now, we turn to the comparison of simulated profiles with observations. In the two innermost bins ($r$ < 0.1 pc), we find that \STARFORGE profile is closer to the AFGL 5180 data points. While our profiles are generally lower than AFGL 5180 in the innermost bin, our M300H profile reaches similar values around an early cluster age of $t$ = 1.0 Myr. In addition, the second AFGL 5180 data point is reached by our M300L and M300H models between 1.0 Myr and 3.0 Myr. Moreover, we find that the AFGL 5180 profile remains relatively flat, with our models being a better match in the power-law index ($\alpha$) (see Table~\ref{tab:powerlawfits}). The \STARFORGE profiles are considerably steeper so the outer AFGL 5180 data points ($r$ > 0.1 pc) are much higher than the corresponding \STARFORGE point. The closest overall match of our M300 models with the AFGL 5180 observations occurs at a cluster age between 1.0 and 2.0 Myr. This is in good agreement with the reported age between 1.0 and 3.0 Myr for the AFGL 5180 region \citep{Devine2008AJ....135.2095D, Vasyunina2010PhDT........23V, Maity2023MNRAS.523.5388M}. Even though we cannot constrain the shape of the profile for G19.88-0.53 with only one point, we note that their measurement aligns well with both our and \STARFORGE models within the range of one standard deviation.

Finally, the third row of Figure~\ref{fig:bestfitmodelNstarprofile} helps us quantify how the power-law index ($\alpha$) of the \avgNstar profile changes with the final stellar mass (\finalstellarmass) of the model. We find that the index becomes less negative with rising \finalstellarmass for our TCCA models, in other words, the profiles are flatter for M3000 and M30000 models than M300 models. The AFGL 5180 profile has a flatter power-law index ($\alpha = -0.690$), which is closer to TCCA models ($\alpha \sim $ -0.7 to -0.8), as observed in the first two rows of Figure~\ref{fig:bestfitmodelNstarprofile} and Table~\ref{tab:powerlawfits}. The \STARFORGE profiles with indices $\alpha \sim -1.3$ are much steeper than both AFGL 5180 and TCCA models.

\begin{table*}
\centering
\caption{Calculated values of constants $A$ and $\alpha$ for power-law fits of the form $\avgNstar (r) = A r^{\alpha}$ to projected stellar density profiles from our simulations (TCCA), \STARFORGE simulations (SF) and AFGL 5180 observations.}

\label{tab:powerlawfits}
\begin{tabular}{ccrrrrrr}
& & \multicolumn{2}{c}{$\minmasscomp \geq 0.01 \msun$} & \multicolumn{2}{c}{$\minmasscomp \geq 0.1 \msun$} & \multicolumn{2}{c}{$\minmasscomp \geq 1.0 \msun$} \\ 
\cline{3-4} \cline{5-6} \cline{7-8} 
Set & \avgNstar Profile & $A$ & $\alpha$ & $A$ & $\alpha$ & $A$ & $\alpha$ \\ \hline

AFGL 5180 & & -- & -- & 147.84 & -0.690 & -- & -- \\ \cline{1-8} 

 & 10 \msun & 27.71 & -1.166 & 19.50 & -1.228 & 1.59 & -1.531  \\
\STARFORGE (SF) & 15 \msun & 25.40 & -1.214  & 18.30 & -1.269 & 1.24 & -1.680  \\
 & 20 \msun & 22.20 & -1.256 & 15.74 & -1.321 & 0.84 & -1.815 \\ \cline{1-8}
 
 & 1.0 Myr & 170.75 & -0.727 &  98.79 & -0.729 &  8.59 & -0.828 \\ 
 & 2.0 Myr & 240.92 & -0.617 & 134.09 & -0.638 & 10.24 & -0.907 \\ 
TCCA M300L & 3.0 Myr & 182.02 & -0.531 & 98.13 & -0.581 & 4.18 & -1.141 \\ 
 & 4.0 Myr & 93.08  & -0.655 &  45.05 & -0.771 &  2.75 & -1.241 \\ 
 & 5.0 Myr & 74.23  & -0.651 &  36.23 & -0.763 &  2.44 & -1.262 \\ \cline{1-8}
 
 & 1.0 Myr & 102.01 & -0.971 & 49.14 & -1.063& 2.16 & -1.552 \\ 
 & 2.0 Myr & 111.47 & -0.738 & 56.69 & -0.814 & 2.83 & -1.341 \\ 
TCCA M300H & 3.0 Myr & 92.40  & -0.679 & 40.70 & -0.804 & 1.57 & -1.448 \\ 
 & 4.0 Myr &  63.19 & -0.763 & 27.21 & -0.906 & 1.36 & -1.480 \\ 
 & 5.0 Myr &  61.91 & -0.710 & 30.30 & -0.822 & 1.70 & -1.399 \\  \cline{1-8}

  & 1.0 Myr & 686.54 & 0.035 & 274.40 & -0.171 & 0.12 & -2.133\\ 
 & 2.0 Myr & no fit & no fit & no fit & no fit & no fit & no fit \\ 
TCCA M3000L & 3.0 Myr & 901.88 & -0.342 & 446.85 & -0.398 & 12.74 & -0.952 \\ 
 & 4.0 Myr &  996.12 & -0.335 & 658.79 & -0.248 & 74.39 & -0.286 \\ 
 & 5.0 Myr &  1221.89 & -0.283 & 652.82 & -0.331 & 47.76 & -0.530 \\  \cline{1-8}

  & 1.0 Myr & 2058.91 & -0.699 & 1084.18 & -0.720 & 76.96 & -0.943 \\ 
 & 2.0 Myr & 1876.13 & -0.470 & 1015.06 & -0.501 & 82.34 & -0.755 \\ 
TCCA M3000H & 3.0 Myr & 1990.77 & -0.313 & 1107.73 & -0.339 & 70.46 & -0.721 \\ 
 & 4.0 Myr &  1608.57 & -0.288 & 903.71 & -0.298 & 77.88 & -0.549 \\ 
 & 5.0 Myr &  1166.46 & -0.255 & 683.29 & -0.275 & 65.30 & -0.528 \\  \cline{1-8}

  & 1.0 Myr & no fit & no fit & no fit & no fit & no fit & no fit \\ 
 & 2.0 Myr & 694.05 & -0.337 & 443.44 & -0.300 & 67.01 & -0.159 \\ 
TCCA M30000L & 3.0 Myr & 1209.10 & -0.304 & 622.00 & -0.315 & 34.48 & -0.668 \\ 
 & 4.0 Myr &  2016.21 & -0.120 & 1024.41 & -0.175 & 108.40 & -0.225 \\ 
 & 5.0 Myr &  2303.86 & -0.147 & 1131.54 & -0.208 & 117.05 & -0.293 \\  \cline{1-8}

  & 1.0 Myr & 8769.36 & -0.401 & 3833.51 & -0.542 & 334.76 & -0.621 \\ 
 & 2.0 Myr & 13799.42 & -0.186 & 7398.29 & -0.209 & 847.74 & -0.241 \\ 
TCCA M30000H & 3.0 Myr & 11282.22 & -0.119 & 6071.45 & -0.141 & 736.41 & -0.190 \\ 
 & 4.0 Myr &  10258.03 & -0.124 & 5618.23 & -0.137 & 636.06 & -0.199\\ 
 & 5.0 Myr &  8009.56 & -0.076 & 4327.21 & -0.104 & 702.76 & 0.021 \\  \cline{1-8}
 
\end{tabular}
\end{table*}

\subsubsection{Comparison of companion properties}
\label{subsubsec:results-observations-companions}

\begin{figure*}
    \centering
    \includegraphics[width=0.9\textwidth]{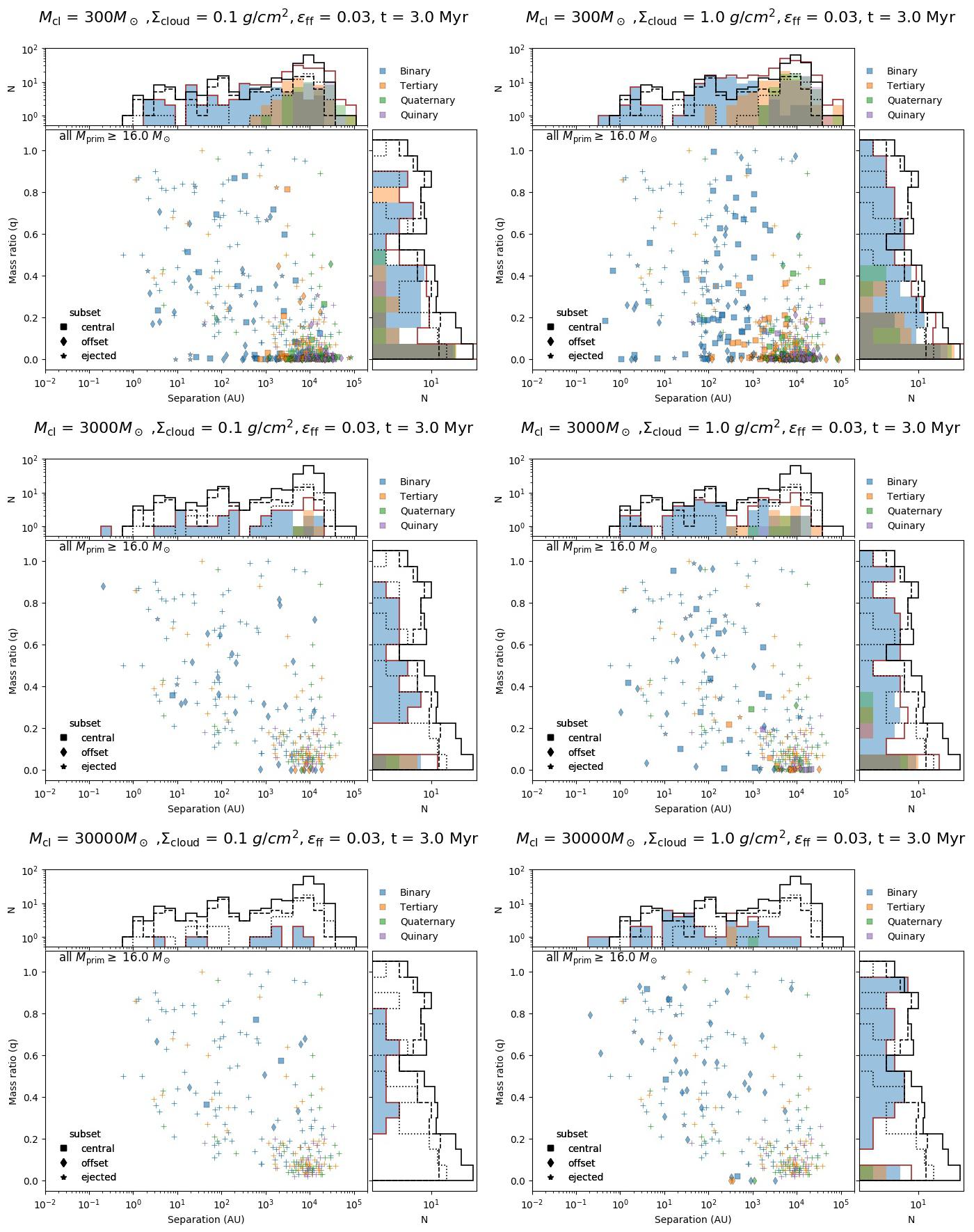}
    \caption{Each panel shows the distribution of the separation and mass-ratios (with respect to primary mass) of bound companions around all massive stars with masses $\geq 16.0 \: \msun$ in our fiducial (\sfeff = 0.03) simulations. From top to bottom, the rows show M300, M3000, and M30000 models, respectively. The left and right columns show low ($\Sigmacloud = 0.1\: \rm g\: cm^{-2}$) and high ($\Sigma_{\rm cloud}= 1.0\: \rm g\: cm^{-2}$) cloud mass surface density cases, respectively. In the scatter plot in each panel, the filled symbols represent our simulated companions, whereas the plus symbols denote observed companions around O-stars from the \texttt{SMASH+} survey. The square, diamond and star markers represent companions around central, offset, and ejected primaries, respectively, in our simulations. The companions are color-coded based on their multiplicity order (binary, triple, quaternary and quinary companions). In the histograms for separation (top panel) and mass-ratio (right panel), the black histogram represents the observed companions, while the red histogram represents the simulated companions. We further plot separate color-coded filled histograms for different orders of simulated companions. For comparison, the binary and tertiary subsets of observed companions are denoted by the dashed and dotted black histograms, respectively. We find that some observed massive stars have close triple companions, which might need to be formed primordially.}
    \label{fig:separation-massratio_allprimaries}
\end{figure*}

Figure~\ref{fig:separation-massratio_allprimaries} shows the separation and mass-ratio of bound companions around all massive stars with masses $ \geq 16.0\: \msun$ in our simulations. We compare them with observed companions around O-stars from the Southern Massive Stars at High Angular Resolution survey \citep[\texttt{SMASH+;}][]{Tramper2025arXiv250918431T}. This high angular resolution interferometric survey targeted nearby O-stars in the Southern hemisphere ($\delta < 0^{\circ}$) and identified companions around them. In addition to the comparison of full set of companions, we compare the different orders of companions separately, focusing on the binary and tertiary companions. We found M300L and M300H models to be the most comparable to their observations, because massive stars in higher \Mcl models have fewer companions (mostly singles or binaries, hence the smaller \avgNcomp).  

The survey had reported a relative scarcity of massive companions at large separations beyond 100 au, with few companions above $q$ > 0.8 at such distances, which were mostly upper limits \citep{Tramper2025arXiv250918431T}. We observe a similar trend in our simulated companions in the M300 models, with large-distance companions having small mass ratios ($ q < 0.4$) beyond 1000 au. On the other hand, the \texttt{SMASH+} survey found close binary and tertiary companions within 10 au of some of these O-stars. In our simulations, we reproduce close binaries however, we lack close triple companions. Our closest triple companions are at separations of 100s of au. We discuss the implications of this lack of close triple companions in section~\ref{subsec:discussion-companion properties}. 

For M300L and M300H models, we ran two-sample Kolmogorov-Smirnov (K-S) test to check whether the similarity of the simulated and observed companion distributions for both separations and mass-ratios. The results of the K-S test are presented in Table~\ref{tab:kstest}. We found that both M300L and M300H models were similar to the observed triple separations while only M300L was similar to the observed binary separations. For mass-ratios however, we did not find similarity between simulated and observed companions in these models.

\begin{table}
\centering
\caption{Results from the two-sample Kolmogorov-Smirnov (K-S) test between the separation and mass-ratio distributions for observed and simulated companions. The null hypothesis ($H_0$) states that both distributions are drawn from the same population. $D$ is the K-S distance between the two distributions being tested for similarity. Each row shows the simulated distribution, compared with the corresponding distribution from \texttt{SMASH+} observations.}

\label{tab:kstest}
\begin{tabular}{ccrrr}

Set & Property & $D$ & $p$-value & $H_0$ \\ \hline

M300L & binary separations & 0.183 & 0.097 & True \\ 
 & triple separations & 0.203  & 0.190 & True \\
 & binary mass-ratios & 0.426  & $< 0.001$ & False \\ 
 & triple mass-ratios & 0.756  & $< 0.001$ & False \\ \cline{1-5}
 
M300H & binary separations & 0.325 & $< 0.001$ & False \\ 
 & triple separations & 0.191  & 0.133 & True \\
 & binary mass-ratios & 0.267  & $< 0.001$ & False \\ 
 & triple mass-ratios & 0.738  & $< 0.001$ & False \\ \cline{1-5}
 
\end{tabular}
\end{table}

\section{Discussion}
\label{sec:discussion}

In section~\ref{subsec:discussion-clusterenvironment}, we discuss the effect of the star cluster environment on the bound systems around massive stars. In section~\ref{subsec:discussion-densityprofile-theories}, we discuss how projected stellar density profiles reflect the clustering of companions predicted by different massive star formation models. In section~\ref{subsec:discussion-companion properties}, we discuss the comparison of our bound companions with observed comparisons around O-stars. In section~\ref{subsec:discussion-modelcaveats}, we mention the caveats of our modeling and the scope for future improvements. 

\subsection{Massive stars in the environment of a star cluster}
\label{subsec:discussion-clusterenvironment}

We found that stars in the central regions of the star cluster have more bound companions than stars of similar mass in the periphery or those that had been ejected from the star cluster. This indicates that, in addition to its mass, the location of a massive star also determines the extent of bound systems that can form around it. This is related to the prevailing number density of stars that could be captured by the massive star. The crowded environment in the center of the star cluster enables massive stars to gather more companions during their formation and/or migration during dynamical mass-segregation. However, this is also true for encounters that could disrupt the bound system. We found that this process also depends on the velocity dispersion of the bound cluster. In particular, we found that massive stars in central regions of star clusters with lower values of stellar velocity dispersion gathered more companions than those in clusters with a higher stellar velocity dispersion. This is because the lower velocity dispersion enhances the gravitational focusing experienced by a potential companion star during the encounter with the massive star and its bound multiple system. This means that captures are favored over disruptions in lower velocity dispersion environments. In higher velocity dispersion environments, the captures become harder because encounters are more energetic on average. Therefore, the extent of bound systems is determined by the stellar velocity dispersion and number density of the star cluster environment. 

The time evolution of stellar velocity dispersion and number density of systems in the bound cluster means that the bound systems around massive stars evolve with time. In early stages of formation when new stars are being formed and number densities are high, the bound systems build up. However, once the star cluster expands after gas removal and the number densities fall, bound systems lose members due to dynamical encounters and/or stripping of weakly bound members by the rest of the star cluster. These stripped companions would become part of the field star population that remains bound to the star cluster.

During our analysis, we identified ejected stars that were unbound from their star cluster. They comprised around 1-30\% of the primary massive star population at $t$ = 3.0 Myr, depending on the model. These stars could have become unbound during the phase of cluster expansion after loss of gas potential. In addition, some stars might have been ejected through three-body encounters in the crowded region of the cluster or during the chaotic breakup of non-hierarchical multiple systems \citep{Poveda1967BOTT....4...86P, Anosova1986Ap&SS.124..217A,Reipurth2010ApJ...725L..56R, Reipurth2026arXiv260206465R}, in the dynamical ejection scenario (DES). Some stars might have received supernova kicks at the end of stellar evolution, either as single stars or part of a binary system, in the binary supernova (BSS) scenario \citep{Blaauw1961BAN....15..265B, Dincel2025arXiv251114686D}. \citet{CarreteroCastrillo2025arXiv251021577C} conducted a study of galactic runaway O-stars studying the signatures of the DES and BSS scenarios on the runaway speed and rotation of such runaway massive stars.

\subsection{Projected stellar density profiles and theories of massive star formation}
\label{subsec:discussion-densityprofile-theories}

Our TCCA models can be considered to represent a limiting case of the Turbulent Core Accretion \citep{McKee2003ApJ...585..850M} model of massive star formation because massive stars are born randomly across the star cluster. They are not necessarily formed in regions of existing high stellar density, in which case they would form with many low-mass companions also forming around them and competing for gas accretion, as in models of Competitive Accretion \citep{Bonnell2001MNRAS.323..785B,Grudic2022MNRAS.512..216G}. We found that the TCCA model, which develops enhanced stellar multiplicity and clustering around massive stars dynamically after individual stars have formed, and Competitive Accretion models that have these features primordially as a crucial part of the massive star formation process, predict different shapes for projected stellar density profiles around massive stars. The Competitive Accretion models have more substructure due to the close low-mass companions. Hence, their projected stellar density profiles are steeper in comparison with the TCCA models, where star formation is more homogeneous. 

In addition, we noted the time evolution for projected stellar density profiles around massive stars in young forming star clusters. In TCCA, the projected stellar density profile builds up quickly over time as more stars are formed around the massive star. However, once the star cluster is formed and gas is dissipated away, the projected stellar density falls as the cluster expands. In star clusters that form quicker (for instance, higher \sfeff models), the projected stellar density profiles can build up and then fall to a baseline level quickly within the first few Myrs. This is true for Competitive Accretion as well, however the massive star takes time to grow in mass and gather enough companions, so the projected stellar density profile is still rising in the first few Myrs in those models. 

We also compared power-law fits between simulated and observed profiles from AFGL 5180 and G19.88-0.53 star-forming regions. We found that G19.88-0.53 was in agreement with both TCCA and Competitive Accretion models. For AFGL 5180, we found that the Competitive Accretion models fitted the observed profile in the two innermost bins ($r$ < 0.1 pc), however they were much steeper and fell lower at larger distances. In contrast, the TCCA models were relatively flat with the power-law indices being closer to that of AFGL 5180 profile. The closest fit of the TCCA models (M300L and M300H) to AFGL 5180 observations occurred at $t \lesssim$ 3.0 Myr. 

However, we have only compared with two high-mass star-forming regions here; observations of more such regions will be necessary to generalize our findings. In addition, the current observations do not measure the profile at smaller scales ($r < 0.01 \rm pc$) due to resolution limits. Our work aimed to test whether, given an observed companion distribution, we could distinguish profiles resulting from different high-mass star formation scenarios. Higher-resolution observations, which can probe the small-scale distribution of companions, will be necessary to confirm the dominant mechanism. It is possible that different scenarios could contribute to the high-mass stellar population in different parts of the profile. If we find that both mechanisms operate together in the same star-forming region, this may indicate a scale-dependent behaviour: small-scale clustering around massive stars could be shaped by localized fragmentation and accretion processes, while the larger-scale radial structure reflects the overall mass distribution. Alternatively, the flatter observed slopes may result from partial dynamical mixing that smoothens any primordial substructure. To explore this, simulations of TCCA and Competitive Accretion tailored for those specific star-forming regions would help to fairly ascertain the relative contributions of these mechanisms to the formation of the overall massive star population. Ideally, these simulations will have enough independent realizations to provide a statistically large sample of massive stars at cluster ages comparable to young star-forming regions.

\subsection{Companion properties}
\label{subsec:discussion-companion properties}

We have 50\% primordial binaries, but no primordial triples and/or multiple systems. However, we found that massive stars gathered many lower-mass companions via dynamical interactions after they are formed in the simulation. We compared the projected separations and mass ratios of companions around massive stars above 16 \msun in our simulations with observed companions around O-stars in the \texttt{SMASH+} survey. We found a similar drop-off in equal-mass companions beyond a certain distance. They noted a decline after 100 au, whereas we found a decline after 1000 au. The fact that we find massive companions further out is probably because our massive stars are part of a forming star cluster with higher stellar densities around them, whereas \texttt{SMASH+} survey includes field stars. So, some dynamical processing may have stripped the companions beyond 100 au around these stars after star cluster formation and expansion was complete. 

In addition, we find close binary companions to massive stars within 10 au of our massive stars, some of which have formed dynamically. \cite{Chon2026arXiv260106251C} used their star cluster simulations to note that such close binary systems can form through the hardening of the binary due to circumbinary disc-driven migration. On the other hand, we lack close triple companions in our simulations. We infer that these close triple companions around observed O-stars must have been formed primordially. These primordial companions could be formed via fragmentation in a core or disk, with the remnant gas possibly driving migration to tighten the orbits of the bound triple system. 

\subsection{Model caveats and future work}
\label{subsec:discussion-modelcaveats}

We have studied the clustering of companions around massive stars in star clusters by post-processing simulations of gradual star cluster formation. These are pure $N$-body simulations with a semi-analytical gas potential. This was necessary because our focus was on studying $N$-body dynamics of stars using multiple realizations for star clusters that were forming at different timescales. Pure hydrodynamical simulations would be too computationally expensive for this purpose and miss the detailed $N$-body dynamics. Recent works using the \texttt{TORCH} framework have made progress in this regard \citep{Wall2019ApJ...887...62W, CournoyerCloutier2024ApJ...977..203C}, even though these coupled $N$-body and hydro simulations are still computationally expensive in general. Nonetheless, capturing the dynamical formation of multiple systems in an evolving gas potential of a forming star cluster will need self-consistent modeling of stellar dynamics, gas hydrodynamics, and stellar feedback, which we aim to include in future work.

Such self-consistent simulations are needed to study the survival of young star clusters as they emerge from their parent gas clouds. During this period, these young clusters can dissolve rapidly due to the change in gravitational potential caused by gas expulsion through effects of stellar feedback \citep[e.g.,][]{Lada2003ARA&A..41...57L}. However, they can also be dissolved by the disruptive effects of continuous tidal shocks generated during interactions of the young star cluster with the surrounding dense gaseous environment \citep[`cruel cradle effect';][]{Elmegreen2010ApJ...712..604E, Kruijssen2011MNRAS.414.1339K, Kruijssen2012MNRAS.419..841K}. In addition, star cluster formation could occur through the merging of fractal substructures formed during the turbulent fragmentation of a single parent molecular cloud \citep{Bonnell2003MNRAS.343..413B, Karam2022MNRAS.513.6095K}, or those formed during cloud-cloud collisions \citep{Wu2017ApJ...841...88W}. These simulations will allow us to study how multiple systems inside such rapidly evolving star clusters react to the changing stellar number density and velocity dispersion of those environments.

Although our simulations did not include primordial mass segregation, some level of mass segregation is developed dynamically during the formation phase, especially in lower-mass clusters \citep[see ][]{Farias2023MNRAS.523.2083F}. We found that this migration of massive stars to regions of higher number densities in the center contributed partially to the formation of multiple systems. Therefore, if massive stars do form preferentially in the crowded regions \citep[for e.g.][]{Bonnell2001MNRAS.323..785B}, it is possible that the number of multiples forming in such environment are higher that what we report here under the same global conditions. Finally, we did not have any primordial triples and/or multiple systems in our simulations. However, observations indicate a population of close triple companions besides close binaries. We aim to include these primordial triples in our future simulations. This will allow us to explore the evolution of young star clusters with simultaneous presence of mass segregation and primordial higher-order multiplicity, similar to recent work by \citet{Amiri2026arXiv260604509A}.

\section{Conclusions}

In this paper, we studied the clustering of bound and unbound companions around massive stars inside gradually forming star clusters via ``Turbulent Clump Core Accretion'' (TCCA) by post-processing existing $N-$body simulations. In these simulations, star clusters are formed in different formation timescales (\tform) parameterized by star formation efficiency per freefall time (\sfeff), mass of star-forming clump (\Mcl) and mass surface densities (\Sigmacloud) of the parent cloud. 

We found that massive stars in the center of bound cluster gather more companions than massive stars that have been ejected from the cluster. We found that ejected stars are mostly single or in binaries. The number of companions is greater for massive stars in star clusters with lower stellar velocity dispersion and higher stellar densities.  When we compared the clustering of companions around our massive stars with projected stellar density profiles, we found a good match to YSO profiles in the AFGL 5180 and G19.88-0.53 star-forming regions. Finally, we noted a lack of close triple companions around our simulated O-stars when compared to companions around O-stars in the \texttt{SMASH+} survey. Since these companions were not captured dynamically, they are likely formed via primordial routes like core or disk fragmentation.   

In the future, we aim to extend our gradual star cluster formation models with hydrodynamical coupling and an accurate treatment of primordial triples. We are also currently working on extending the analysis of dynamically gathered companions around primary stars across the full stellar mass spectrum. 

\section*{Acknowledgements}
We thank the anonymous reviewer for their comments and constructive feedback that greatly improved the manuscript.
We thank Mike Grudi\'{c} and Hugues Sana for helpful discussions.
AG acknowledges invaluable support from the Chalmers Astrophysics and Space Sciences Summer (CASSUM) research program. JCT acknowledges support from ERC Advanced Grant 788829 (MSTAR), NSF grant AST-2009674, and funding from the Virginia Institute for Theoretical Astrophysics (VITA), supported by the College and Graduate School of Arts and Sciences at the University of Virginia. 
JPF acknowledges support from the NASA grant 80NSSC20K0507 and NSF Career award 1748571.
The simulations presented in this work were performed on the Origins computing cluster at Chalmers. The authors acknowledge Research Computing at The University of Virginia for providing computational resources and technical support that have contributed to the results reported within this publication. URL: https://rc.virginia.edu. This research has made use of NASA’s Astrophysics Data System.  

\section*{Data Availability}
The data underlying this article will be shared on reasonable request to the corresponding author.



\bibliographystyle{mnras}
\bibliography{references} 

\begin{thebibliography}{}
\makeatletter
\relax
\def\mn@urlcharsother{\let\do\@makeother \do\$\do\&\do\#\do\^\do\_\do\%\do\~}
\def\mn@doi{\begingroup\mn@urlcharsother \@ifnextchar [ {\mn@doi@} {\mn@doi@[]}}
\def\mn@doi@[#1]#2{\def\@tempa{#1}\ifx\@tempa\@empty \href {http://dx.doi.org/#2} {doi:#2}\else \href {http://dx.doi.org/#2} {#1}\fi \endgroup}
\def\mn@eprint#1#2{\mn@eprint@#1:#2::\@nil}
\def\mn@eprint@arXiv#1{\href {http://arxiv.org/abs/#1} {{\tt arXiv:#1}}}
\def\mn@eprint@dblp#1{\href {http://dblp.uni-trier.de/rec/bibtex/#1.xml} {dblp:#1}}
\def\mn@eprint@#1:#2:#3:#4\@nil{\def\@tempa {#1}\def\@tempb {#2}\def\@tempc {#3}\ifx \@tempc \@empty \let \@tempc \@tempb \let \@tempb \@tempa \fi \ifx \@tempb \@empty \def\@tempb {arXiv}\fi \@ifundefined {mn@eprint@\@tempb}{\@tempb:\@tempc}{\expandafter \expandafter \csname mn@eprint@\@tempb\endcsname \expandafter{\@tempc}}}

\bibitem[\protect\citeauthoryear{{Alcal{\'a}} et~al.,}{{Alcal{\'a}} et~al.}{2014}]{Alcala2014A&A...561A...2A}
{Alcal{\'a}} J.~M.,  et~al., 2014, \mn@doi [\aap] {10.1051/0004-6361/201322254}, \href {https://ui.adsabs.harvard.edu/abs/2014A&A...561A...2A} {561, A2}

\bibitem[\protect\citeauthoryear{{Allison}, {Goodwin}, {Parker}, {de Grijs}, {Portegies Zwart}  \& {Kouwenhoven}}{{Allison} et~al.}{2009}]{Allison2009ApJ...700L..99A}
{Allison} R.~J.,  {Goodwin} S.~P.,  {Parker} R.~J.,  {de Grijs} R.,  {Portegies Zwart} S.~F.,   {Kouwenhoven} M.~B.~N.,  2009, \mn@doi [\apjl] {10.1088/0004-637X/700/2/L99}, \href {https://ui.adsabs.harvard.edu/abs/2009ApJ...700L..99A} {700, L99}

\bibitem[\protect\citeauthoryear{{Amiri}, {Flammini Dotti}, {Pang}, {Kamlah}, {Berczik}, {Shukirgaliyev}  \& {Spurzem}}{{Amiri} et~al.}{2026}]{Amiri2026arXiv260604509A}
{Amiri} V.,  {Flammini Dotti} F.,  {Pang} X.,  {Kamlah} A.~W.~H.,  {Berczik} P.,  {Shukirgaliyev} B.,   {Spurzem} R.,  2026, \mn@doi [arXiv e-prints] {10.48550/arXiv.2606.04509}, \href {https://ui.adsabs.harvard.edu/abs/2026arXiv260604509A} {p. arXiv:2606.04509}

\bibitem[\protect\citeauthoryear{{Anosova}}{{Anosova}}{1986}]{Anosova1986Ap&SS.124..217A}
{Anosova} J.~P.,  1986, \mn@doi [\apss] {10.1007/BF00656037}, \href {https://ui.adsabs.harvard.edu/abs/1986Ap&SS.124..217A} {124, 217}

\bibitem[\protect\citeauthoryear{{Appel} et~al.,}{{Appel} et~al.}{2026}]{Appel2026AJ....172...12A}
{Appel} S.~M.,  et~al., 2026, \mn@doi [\aj] {10.3847/1538-3881/ae633f}, \href {https://ui.adsabs.harvard.edu/abs/2026AJ....172...12A} {172, 12}

\bibitem[\protect\citeauthoryear{{Bate} \& {Bonnell}}{{Bate} \& {Bonnell}}{1997}]{BateBonnell1997MNRAS.285...33B}
{Bate} M.~R.,  {Bonnell} I.~A.,  1997, \mn@doi [\mnras] {10.1093/mnras/285.1.33}, \href {https://ui.adsabs.harvard.edu/abs/1997MNRAS.285...33B} {285, 33}

\bibitem[\protect\citeauthoryear{{Beuther}, {Kuiper}  \& {Tafalla}}{{Beuther} et~al.}{2025}]{Beuther2025ARA&A..63....1B}
{Beuther} H.,  {Kuiper} R.,   {Tafalla} M.,  2025, \mn@doi [\araa] {10.1146/annurev-astro-013125-122023}, \href {https://ui.adsabs.harvard.edu/abs/2025ARA&A..63....1B} {63, 1}

\bibitem[\protect\citeauthoryear{{Blaauw}}{{Blaauw}}{1961}]{Blaauw1961BAN....15..265B}
{Blaauw} A.,  1961, \bain, \href {https://ui.adsabs.harvard.edu/abs/1961BAN....15..265B} {15, 265}

\bibitem[\protect\citeauthoryear{{Bonnell}, {Bate}, {Clarke}  \& {Pringle}}{{Bonnell} et~al.}{2001}]{Bonnell2001MNRAS.323..785B}
{Bonnell} I.~A.,  {Bate} M.~R.,  {Clarke} C.~J.,   {Pringle} J.~E.,  2001, \mn@doi [\mnras] {10.1046/j.1365-8711.2001.04270.x}, \href {https://ui.adsabs.harvard.edu/abs/2001MNRAS.323..785B} {323, 785}

\bibitem[\protect\citeauthoryear{{Bonnell}, {Bate}  \& {Vine}}{{Bonnell} et~al.}{2003}]{Bonnell2003MNRAS.343..413B}
{Bonnell} I.~A.,  {Bate} M.~R.,   {Vine} S.~G.,  2003, \mn@doi [\mnras] {10.1046/j.1365-8711.2003.06687.x}, \href {https://ui.adsabs.harvard.edu/abs/2003MNRAS.343..413B} {343, 413}

\bibitem[\protect\citeauthoryear{{Butler} \& {Tan}}{{Butler} \& {Tan}}{2012}]{Butler2012ApJ...754....5B}
{Butler} M.~J.,  {Tan} J.~C.,  2012, \mn@doi [\apj] {10.1088/0004-637X/754/1/5}, \href {https://ui.adsabs.harvard.edu/abs/2012ApJ...754....5B} {754, 5}

\bibitem[\protect\citeauthoryear{{Butler}, {Tan}, {Teyssier}, {Rosdahl}, {Van Loo}  \& {Nickerson}}{{Butler} et~al.}{2017}]{Butler2017ApJ...841...82B}
{Butler} M.~J.,  {Tan} J.~C.,  {Teyssier} R.,  {Rosdahl} J.,  {Van Loo} S.,   {Nickerson} S.,  2017, \mn@doi [\apj] {10.3847/1538-4357/aa7054}, \href {https://ui.adsabs.harvard.edu/abs/2017ApJ...841...82B} {841, 82}

\bibitem[\protect\citeauthoryear{{Carretero-Castrillo}, {Rib{\'o}}, {Paredes}, {Holgado}, {Mart{\'\i}nez-Sebasti{\'a}n}  \& {Sim{\'o}n-D{\'\i}az}}{{Carretero-Castrillo} et~al.}{2025}]{CarreteroCastrillo2025arXiv251021577C}
{Carretero-Castrillo} M.,  {Rib{\'o}} M.,  {Paredes} J.~M.,  {Holgado} G.,  {Mart{\'\i}nez-Sebasti{\'a}n} C.,   {Sim{\'o}n-D{\'\i}az} S.,  2025, \mn@doi [arXiv e-prints] {10.48550/arXiv.2510.21577}, \href {https://ui.adsabs.harvard.edu/abs/2025arXiv251021577C} {p. arXiv:2510.21577}

\bibitem[\protect\citeauthoryear{{Chabrier}}{{Chabrier}}{2003}]{Chabrier2003PASP..115..763C}
{Chabrier} G.,  2003, \mn@doi [\pasp] {10.1086/376392}, \href {https://ui.adsabs.harvard.edu/abs/2003PASP..115..763C} {115, 763}

\bibitem[\protect\citeauthoryear{{Chandrasekhar}}{{Chandrasekhar}}{1943}]{Chandrasekhar1943ApJ....97..255C}
{Chandrasekhar} S.,  1943, \mn@doi [\apj] {10.1086/144517}, \href {https://ui.adsabs.harvard.edu/abs/1943ApJ....97..255C} {97, 255}

\bibitem[\protect\citeauthoryear{{Chon} \& {Vigna-G{\'o}mez}}{{Chon} \& {Vigna-G{\'o}mez}}{2026}]{Chon2026arXiv260106251C}
{Chon} S.,  {Vigna-G{\'o}mez} A.,  2026, \mn@doi [arXiv e-prints] {10.48550/arXiv.2601.06251}, \href {https://ui.adsabs.harvard.edu/abs/2026arXiv260106251C} {p. arXiv:2601.06251}

\bibitem[\protect\citeauthoryear{{Costa Silva} et~al.,}{{Costa Silva} et~al.}{2022}]{CostaSilva2022A&A...659A..23C}
{Costa Silva} A.~R.,  et~al., 2022, \mn@doi [\aap] {10.1051/0004-6361/202142412}, \href {https://ui.adsabs.harvard.edu/abs/2022A&A...659A..23C} {659, A23}

\bibitem[\protect\citeauthoryear{{Cournoyer-Cloutier} et~al.,}{{Cournoyer-Cloutier} et~al.}{2024}]{CournoyerCloutier2024ApJ...977..203C}
{Cournoyer-Cloutier} C.,  et~al., 2024, \mn@doi [\apj] {10.3847/1538-4357/ad90b3}, \href {https://ui.adsabs.harvard.edu/abs/2024ApJ...977..203C} {977, 203}

\bibitem[\protect\citeauthoryear{{Crowe} et~al.,}{{Crowe} et~al.}{2024}]{Crowe2024A&A...682A...2C}
{Crowe} S.,  et~al., 2024, \mn@doi [\aap] {10.1051/0004-6361/202348094}, \href {https://ui.adsabs.harvard.edu/abs/2024A&A...682A...2C} {682, A2}

\bibitem[\protect\citeauthoryear{{Da Rio}, {Robberto}, {Soderblom}, {Panagia}, {Hillenbrand}, {Palla}  \& {Stassun}}{{Da Rio} et~al.}{2010}]{DaRio2010ApJ...722.1092D}
{Da Rio} N.,  {Robberto} M.,  {Soderblom} D.~R.,  {Panagia} N.,  {Hillenbrand} L.~A.,  {Palla} F.,   {Stassun} K.~G.,  2010, \mn@doi [\apj] {10.1088/0004-637X/722/2/1092}, \href {https://ui.adsabs.harvard.edu/abs/2010ApJ...722.1092D} {722, 1092}

\bibitem[\protect\citeauthoryear{{Dale}, {Ercolano}  \& {Bonnell}}{{Dale} et~al.}{2012}]{Dale2012MNRAS.424..377D}
{Dale} J.~E.,  {Ercolano} B.,   {Bonnell} I.~A.,  2012, \mn@doi [\mnras] {10.1111/j.1365-2966.2012.21205.x}, \href {https://ui.adsabs.harvard.edu/abs/2012MNRAS.424..377D} {424, 377}

\bibitem[\protect\citeauthoryear{{Dale}, {Ngoumou}, {Ercolano}  \& {Bonnell}}{{Dale} et~al.}{2014}]{Dale2014MNRAS.442..694D}
{Dale} J.~E.,  {Ngoumou} J.,  {Ercolano} B.,   {Bonnell} I.~A.,  2014, \mn@doi [\mnras] {10.1093/mnras/stu816}, \href {https://ui.adsabs.harvard.edu/abs/2014MNRAS.442..694D} {442, 694}

\bibitem[\protect\citeauthoryear{{Devine}, {Churchwell}, {Indebetouw}, {Watson}  \& {Crawford}}{{Devine} et~al.}{2008}]{Devine2008AJ....135.2095D}
{Devine} K.~E.,  {Churchwell} E.~B.,  {Indebetouw} R.,  {Watson} C.,   {Crawford} S.~M.,  2008, \mn@doi [\aj] {10.1088/0004-6256/135/6/2095}, \href {https://ui.adsabs.harvard.edu/abs/2008AJ....135.2095D} {135, 2095}

\bibitem[\protect\citeauthoryear{{Din{\c{c}}el} et~al.,}{{Din{\c{c}}el} et~al.}{2025}]{Dincel2025arXiv251114686D}
{Din{\c{c}}el} B.,  et~al., 2025, \mn@doi [arXiv e-prints] {10.48550/arXiv.2511.14686}, \href {https://ui.adsabs.harvard.edu/abs/2025arXiv251114686D} {p. arXiv:2511.14686}

\bibitem[\protect\citeauthoryear{{Dom{\'\i}nguez}, {Fellhauer}, {Bla{\~n}a}, {Farias}  \& {Dabringhausen}}{{Dom{\'\i}nguez} et~al.}{2017}]{Dominguez2017MNRAS.472..465D}
{Dom{\'\i}nguez} R.,  {Fellhauer} M.,  {Bla{\~n}a} M.,  {Farias} J.~P.,   {Dabringhausen} J.,  2017, \mn@doi [\mnras] {10.1093/mnras/stx1883}, \href {https://ui.adsabs.harvard.edu/abs/2017MNRAS.472..465D} {472, 465}

\bibitem[\protect\citeauthoryear{{Elmegreen}}{{Elmegreen}}{2011}]{Elmegreen2011EAS....51...45E}
{Elmegreen} B.~G.,  2011, in {Charbonnel} C.,  {Montmerle} T.,  eds,  EAS Publications Series Vol. 51, EAS Publications Series. EDP, pp 45--58 (\mn@eprint {arXiv} {1101.3112}), \mn@doi{10.1051/eas/1151004}

\bibitem[\protect\citeauthoryear{{Elmegreen} \& {Hunter}}{{Elmegreen} \& {Hunter}}{2010}]{Elmegreen2010ApJ...712..604E}
{Elmegreen} B.~G.,  {Hunter} D.~A.,  2010, \mn@doi [\apj] {10.1088/0004-637X/712/1/604}, \href {https://ui.adsabs.harvard.edu/abs/2010ApJ...712..604E} {712, 604}

\bibitem[\protect\citeauthoryear{{Fajrin}, {Armstrong}, {Tan}, {Farias}  \& {Eyer}}{{Fajrin} et~al.}{2025}]{Fajrin2025MNRAS.537.1320F}
{Fajrin} M.,  {Armstrong} J.~J.,  {Tan} J.~C.,  {Farias} J.~P.,   {Eyer} L.,  2025, \mn@doi [\mnras] {10.1093/mnras/staf074}, \href {https://ui.adsabs.harvard.edu/abs/2025MNRAS.537.1320F} {537, 1320}

\bibitem[\protect\citeauthoryear{{Farias} \& {Tan}}{{Farias} \& {Tan}}{2023}]{Farias2023MNRAS.523.2083F}
{Farias} J.~P.,  {Tan} J.~C.,  2023, \mn@doi [\mnras] {10.1093/mnras/stad1532}, \href {https://ui.adsabs.harvard.edu/abs/2023MNRAS.523.2083F} {523, 2083}

\bibitem[\protect\citeauthoryear{{Farias}, {Tan}  \& {Chatterjee}}{{Farias} et~al.}{2017}]{Farias2017ApJ...838..116F}
{Farias} J.~P.,  {Tan} J.~C.,   {Chatterjee} S.,  2017, \mn@doi [\apj] {10.3847/1538-4357/aa63f6}, \href {https://ui.adsabs.harvard.edu/abs/2017ApJ...838..116F} {838, 116}

\bibitem[\protect\citeauthoryear{{Farias}, {Tan}  \& {Chatterjee}}{{Farias} et~al.}{2019}]{Farias2019MNRAS.483.4999F}
{Farias} J.~P.,  {Tan} J.~C.,   {Chatterjee} S.,  2019, \mn@doi [\mnras] {10.1093/mnras/sty3470}, \href {https://ui.adsabs.harvard.edu/abs/2019MNRAS.483.4999F} {483, 4999}

\bibitem[\protect\citeauthoryear{{Gammie}}{{Gammie}}{2001}]{Gammie2001ApJ...553..174G}
{Gammie} C.~F.,  2001, \mn@doi [\apj] {10.1086/320631}, \href {https://ui.adsabs.harvard.edu/abs/2001ApJ...553..174G} {553, 174}

\bibitem[\protect\citeauthoryear{{Grudi{\'c}}, {Guszejnov}, {Offner}, {Rosen}, {Raju}, {Faucher-Gigu{\`e}re}  \& {Hopkins}}{{Grudi{\'c}} et~al.}{2022}]{Grudic2022MNRAS.512..216G}
{Grudi{\'c}} M.~Y.,  {Guszejnov} D.,  {Offner} S. S.~R.,  {Rosen} A.~L.,  {Raju} A.~N.,  {Faucher-Gigu{\`e}re} C.-A.,   {Hopkins} P.~F.,  2022, \mn@doi [\mnras] {10.1093/mnras/stac526}, \href {https://ui.adsabs.harvard.edu/abs/2022MNRAS.512..216G} {512, 216}

\bibitem[\protect\citeauthoryear{{Guszejnov}, {Hopkins}  \& {Krumholz}}{{Guszejnov} et~al.}{2017}]{Guszejnov2017MNRAS.468.4093G}
{Guszejnov} D.,  {Hopkins} P.~F.,   {Krumholz} M.~R.,  2017, \mn@doi [\mnras] {10.1093/mnras/stx725}, \href {https://ui.adsabs.harvard.edu/abs/2017MNRAS.468.4093G} {468, 4093}

\bibitem[\protect\citeauthoryear{{Heggie}}{{Heggie}}{1975}]{Heggie1975MNRAS.173..729H}
{Heggie} D.~C.,  1975, \mn@doi [\mnras] {10.1093/mnras/173.3.729}, \href {https://ui.adsabs.harvard.edu/abs/1975MNRAS.173..729H} {173, 729}

\bibitem[\protect\citeauthoryear{{Hills}}{{Hills}}{1975}]{Hills1975AJ.....80.1075H}
{Hills} J.~G.,  1975, \mn@doi [\aj] {10.1086/111842}, \href {https://ui.adsabs.harvard.edu/abs/1975AJ.....80.1075H} {80, 1075}

\bibitem[\protect\citeauthoryear{{Inutsuka} \& {Miyama}}{{Inutsuka} \& {Miyama}}{1997}]{Inutsuka1997ApJ...480..681I}
{Inutsuka} S.-i.,  {Miyama} S.~M.,  1997, \mn@doi [\apj] {10.1086/303982}, \href {https://ui.adsabs.harvard.edu/abs/1997ApJ...480..681I} {480, 681}

\bibitem[\protect\citeauthoryear{{Issac}, {Tej}, {Liu}, {Varricatt}, {Vig}, {Ishwara Chandra}, {Schultheis}  \& {Nandakumar}}{{Issac} et~al.}{2020}]{Issac2020MNRAS.497.5454I}
{Issac} N.,  {Tej} A.,  {Liu} T.,  {Varricatt} W.,  {Vig} S.,  {Ishwara Chandra} C.~H.,  {Schultheis} M.,   {Nandakumar} G.,  2020, \mn@doi [\mnras] {10.1093/mnras/staa2301}, \href {https://ui.adsabs.harvard.edu/abs/2020MNRAS.497.5454I} {497, 5454}

\bibitem[\protect\citeauthoryear{{Karam} \& {Sills}}{{Karam} \& {Sills}}{2022}]{Karam2022MNRAS.513.6095K}
{Karam} J.,  {Sills} A.,  2022, \mn@doi [\mnras] {10.1093/mnras/stac1298}, \href {https://ui.adsabs.harvard.edu/abs/2022MNRAS.513.6095K} {513, 6095}

\bibitem[\protect\citeauthoryear{{Kenyon} \& {Hartmann}}{{Kenyon} \& {Hartmann}}{1995}]{Kenyon1995ApJS..101..117K}
{Kenyon} S.~J.,  {Hartmann} L.,  1995, \mn@doi [\apjs] {10.1086/192235}, \href {https://ui.adsabs.harvard.edu/abs/1995ApJS..101..117K} {101, 117}

\bibitem[\protect\citeauthoryear{{Kinoshita}, {Nakamura}  \& {Wu}}{{Kinoshita} et~al.}{2021}]{Kinoshita2021arXiv210805554K}
{Kinoshita} S.~W.,  {Nakamura} F.,   {Wu} B.,  2021, \mn@doi [arXiv e-prints] {10.48550/arXiv.2108.05554}, \href {https://ui.adsabs.harvard.edu/abs/2021arXiv210805554K} {p. arXiv:2108.05554}

\bibitem[\protect\citeauthoryear{{Kirk} et~al.,}{{Kirk} et~al.}{2017}]{Kirk2017ApJ...838..114K}
{Kirk} H.,  et~al., 2017, \mn@doi [\apj] {10.3847/1538-4357/aa63f8}, \href {https://ui.adsabs.harvard.edu/abs/2017ApJ...838..114K} {838, 114}

\bibitem[\protect\citeauthoryear{{Klein}, {Sandford}  \& {Whitaker}}{{Klein} et~al.}{1980}]{Klein1980SSRv...27..275K}
{Klein} R.~I.,  {Sandford} II M.~T.,   {Whitaker} R.~W.,  1980, \mn@doi [\ssr] {10.1007/BF00168309}, \href {https://ui.adsabs.harvard.edu/abs/1980SSRv...27..275K} {27, 275}

\bibitem[\protect\citeauthoryear{{K{\"o}lligan} \& {Kuiper}}{{K{\"o}lligan} \& {Kuiper}}{2018}]{KolliganKuiper2018A&A...620A.182K}
{K{\"o}lligan} A.,  {Kuiper} R.,  2018, \mn@doi [\aap] {10.1051/0004-6361/201833686}, \href {https://ui.adsabs.harvard.edu/abs/2018A&A...620A.182K} {620, A182}

\bibitem[\protect\citeauthoryear{{Kroupa}}{{Kroupa}}{2001}]{Kroupa2001MNRAS.322..231K}
{Kroupa} P.,  2001, \mn@doi [\mnras] {10.1046/j.1365-8711.2001.04022.x}, \href {https://ui.adsabs.harvard.edu/abs/2001MNRAS.322..231K} {322, 231}

\bibitem[\protect\citeauthoryear{{Kruijssen}, {Pelupessy}, {Lamers}, {Portegies Zwart}  \& {Icke}}{{Kruijssen} et~al.}{2011}]{Kruijssen2011MNRAS.414.1339K}
{Kruijssen} J.~M.~D.,  {Pelupessy} F.~I.,  {Lamers} H. J.~G.~L.~M.,  {Portegies Zwart} S.~F.,   {Icke} V.,  2011, \mn@doi [\mnras] {10.1111/j.1365-2966.2011.18467.x}, \href {https://ui.adsabs.harvard.edu/abs/2011MNRAS.414.1339K} {414, 1339}

\bibitem[\protect\citeauthoryear{{Kruijssen}, {Maschberger}, {Moeckel}, {Clarke}, {Bastian}  \& {Bonnell}}{{Kruijssen} et~al.}{2012}]{Kruijssen2012MNRAS.419..841K}
{Kruijssen} J.~M.~D.,  {Maschberger} T.,  {Moeckel} N.,  {Clarke} C.~J.,  {Bastian} N.,   {Bonnell} I.~A.,  2012, \mn@doi [\mnras] {10.1111/j.1365-2966.2011.19748.x}, \href {https://ui.adsabs.harvard.edu/abs/2012MNRAS.419..841K} {419, 841}

\bibitem[\protect\citeauthoryear{{Lada} \& {Lada}}{{Lada} \& {Lada}}{2003}]{Lada2003ARA&A..41...57L}
{Lada} C.~J.,  {Lada} E.~A.,  2003, \mn@doi [\araa] {10.1146/annurev.astro.41.011802.094844}, \href {https://ui.adsabs.harvard.edu/abs/2003ARA&A..41...57L} {41, 57}

\bibitem[\protect\citeauthoryear{{Lewis} et~al.,}{{Lewis} et~al.}{2023}]{Lewis2023ApJ...944..211L}
{Lewis} S.~C.,  et~al., 2023, \mn@doi [\apj] {10.3847/1538-4357/acb0c5}, \href {https://ui.adsabs.harvard.edu/abs/2023ApJ...944..211L} {944, 211}

\bibitem[\protect\citeauthoryear{{Luhman}}{{Luhman}}{2004}]{Luhman2004ApJ...602..816L}
{Luhman} K.~L.,  2004, \mn@doi [\apj] {10.1086/381146}, \href {https://ui.adsabs.harvard.edu/abs/2004ApJ...602..816L} {602, 816}

\bibitem[\protect\citeauthoryear{{Maity}, {Dewangan}, {Bhadari}, {Ojha}, {Chen}  \& {Pandey}}{{Maity} et~al.}{2023}]{Maity2023MNRAS.523.5388M}
{Maity} A.~K.,  {Dewangan} L.~K.,  {Bhadari} N.~K.,  {Ojha} D.~K.,  {Chen} Z.,   {Pandey} R.,  2023, \mn@doi [\mnras] {10.1093/mnras/stad1644}, \href {https://ui.adsabs.harvard.edu/abs/2023MNRAS.523.5388M} {523, 5388}

\bibitem[\protect\citeauthoryear{{Matzner}}{{Matzner}}{2002}]{Matzner2002ApJ...566..302M}
{Matzner} C.~D.,  2002, \mn@doi [\apj] {10.1086/338030}, \href {https://ui.adsabs.harvard.edu/abs/2002ApJ...566..302M} {566, 302}

\bibitem[\protect\citeauthoryear{{McKee} \& {Tan}}{{McKee} \& {Tan}}{2003}]{McKee2003ApJ...585..850M}
{McKee} C.~F.,  {Tan} J.~C.,  2003, \mn@doi [\apj] {10.1086/346149}, \href {https://ui.adsabs.harvard.edu/abs/2003ApJ...585..850M} {585, 850}

\bibitem[\protect\citeauthoryear{{Moe} \& {Di Stefano}}{{Moe} \& {Di Stefano}}{2017}]{MoeStefano2017ApJS..230...15M}
{Moe} M.,  {Di Stefano} R.,  2017, \mn@doi [\apjs] {10.3847/1538-4365/aa6fb6}, \href {https://ui.adsabs.harvard.edu/abs/2017ApJS..230...15M} {230, 15}

\bibitem[\protect\citeauthoryear{{Nomoto}, {Kobayashi}  \& {Tominaga}}{{Nomoto} et~al.}{2013}]{Nomoto2013ARA&A..51..457N}
{Nomoto} K.,  {Kobayashi} C.,   {Tominaga} N.,  2013, \mn@doi [\araa] {10.1146/annurev-astro-082812-140956}, \href {https://ui.adsabs.harvard.edu/abs/2013ARA&A..51..457N} {51, 457}

\bibitem[\protect\citeauthoryear{{Offner}, {Moe}, {Kratter}, {Sadavoy}, {Jensen}  \& {Tobin}}{{Offner} et~al.}{2023}]{offner2023ASPC..534..275O}
{Offner} S.~S.~R.,  {Moe} M.,  {Kratter} K.~M.,  {Sadavoy} S.~I.,  {Jensen} E.~L.~N.,   {Tobin} J.~J.,  2023, in {Inutsuka} S.,  {Aikawa} Y.,  {Muto} T.,  {Tomida} K.,   {Tamura} M.,  eds,  Astronomical Society of the Pacific Conference Series Vol. 534, Protostars and Planets VII. p.~275 (\mn@eprint {arXiv} {2203.10066}), \mn@doi{10.48550/arXiv.2203.10066}

\bibitem[\protect\citeauthoryear{{Pecaut}, {Mamajek}  \& {Bubar}}{{Pecaut} et~al.}{2012}]{Pecaut2012ApJ...746..154P}
{Pecaut} M.~J.,  {Mamajek} E.~E.,   {Bubar} E.~J.,  2012, \mn@doi [\apj] {10.1088/0004-637X/746/2/154}, \href {https://ui.adsabs.harvard.edu/abs/2012ApJ...746..154P} {746, 154}

\bibitem[\protect\citeauthoryear{{Pineda} et~al.,}{{Pineda} et~al.}{2015}]{Pineda2015Natur.518..213P}
{Pineda} J.~E.,  et~al., 2015, \mn@doi [\nat] {10.1038/nature14166}, \href {https://ui.adsabs.harvard.edu/abs/2015Natur.518..213P} {518, 213}

\bibitem[\protect\citeauthoryear{{Poveda}, {Ruiz}  \& {Allen}}{{Poveda} et~al.}{1967}]{Poveda1967BOTT....4...86P}
{Poveda} A.,  {Ruiz} J.,   {Allen} C.,  1967, Boletin de los Observatorios Tonantzintla y Tacubaya, \href {https://ui.adsabs.harvard.edu/abs/1967BOTT....4...86P} {4, 86}

\bibitem[\protect\citeauthoryear{{Preibisch}}{{Preibisch}}{2012}]{Preibisch2012RAA....12....1P}
{Preibisch} T.,  2012, \mn@doi [Research in Astronomy and Astrophysics] {10.1088/1674-4527/12/1/001}, \href {https://ui.adsabs.harvard.edu/abs/2012RAA....12....1P} {12, 1}

\bibitem[\protect\citeauthoryear{{Raghavan} et~al.,}{{Raghavan} et~al.}{2010}]{Raghavan2010ApJS..190....1R}
{Raghavan} D.,  et~al., 2010, \mn@doi [\apjs] {10.1088/0067-0049/190/1/1}, \href {https://ui.adsabs.harvard.edu/abs/2010ApJS..190....1R} {190, 1}

\bibitem[\protect\citeauthoryear{{Reggiani}, {Robberto}, {Da Rio}, {Meyer}, {Soderblom}  \& {Ricci}}{{Reggiani} et~al.}{2011}]{Reggiani2011A&A...534A..83R}
{Reggiani} M.,  {Robberto} M.,  {Da Rio} N.,  {Meyer} M.~R.,  {Soderblom} D.~R.,   {Ricci} L.,  2011, \mn@doi [\aap] {10.1051/0004-6361/201116946}, \href {https://ui.adsabs.harvard.edu/abs/2011A&A...534A..83R} {534, A83}

\bibitem[\protect\citeauthoryear{{Reipurth}, {Mikkola}, {Connelley}  \& {Valtonen}}{{Reipurth} et~al.}{2010}]{Reipurth2010ApJ...725L..56R}
{Reipurth} B.,  {Mikkola} S.,  {Connelley} M.,   {Valtonen} M.,  2010, \mn@doi [\apjl] {10.1088/2041-8205/725/1/L56}, \href {https://ui.adsabs.harvard.edu/abs/2010ApJ...725L..56R} {725, L56}

\bibitem[\protect\citeauthoryear{{Reipurth} et~al.,}{{Reipurth} et~al.}{2026}]{Reipurth2026arXiv260206465R}
{Reipurth} B.,  et~al., 2026, \mn@doi [arXiv e-prints] {10.48550/arXiv.2602.06465}, \href {https://ui.adsabs.harvard.edu/abs/2026arXiv260206465R} {p. arXiv:2602.06465}

\bibitem[\protect\citeauthoryear{{Reynolds} et~al.,}{{Reynolds} et~al.}{2021}]{Reynolds2021ApJ...907L..10R}
{Reynolds} N.~K.,  et~al., 2021, \mn@doi [\apjl] {10.3847/2041-8213/abcc02}, \href {https://ui.adsabs.harvard.edu/abs/2021ApJ...907L..10R} {907, L10}

\bibitem[\protect\citeauthoryear{{Rogers} \& {Pittard}}{{Rogers} \& {Pittard}}{2013}]{RogersPittard2013MNRAS.431.1337R}
{Rogers} H.,  {Pittard} J.~M.,  2013, \mn@doi [\mnras] {10.1093/mnras/stt255}, \href {https://ui.adsabs.harvard.edu/abs/2013MNRAS.431.1337R} {431, 1337}

\bibitem[\protect\citeauthoryear{{Sana} et~al.,}{{Sana} et~al.}{2012}]{Sana2012Sci...337..444S}
{Sana} H.,  et~al., 2012, \mn@doi [Science] {10.1126/science.1223344}, \href {https://ui.adsabs.harvard.edu/abs/2012Sci...337..444S} {337, 444}

\bibitem[\protect\citeauthoryear{{Smith}, {Sijacki}  \& {Shen}}{{Smith} et~al.}{2018}]{Smith2018MNRAS.478..302S}
{Smith} M.~C.,  {Sijacki} D.,   {Shen} S.,  2018, \mn@doi [\mnras] {10.1093/mnras/sty994}, \href {https://ui.adsabs.harvard.edu/abs/2018MNRAS.478..302S} {478, 302}

\bibitem[\protect\citeauthoryear{{Staff}, {Tanaka}  \& {Tan}}{{Staff} et~al.}{2019}]{Staff2019ApJ...882..123S}
{Staff} J.~E.,  {Tanaka} K. E.~I.,   {Tan} J.~C.,  2019, \mn@doi [\apj] {10.3847/1538-4357/ab36b3}, \href {https://ui.adsabs.harvard.edu/abs/2019ApJ...882..123S} {882, 123}

\bibitem[\protect\citeauthoryear{{Staff}, {Tanaka}, {Ramsey}, {Zhang}  \& {Tan}}{{Staff} et~al.}{2023}]{Staff2023ApJ...947...40S}
{Staff} J.~E.,  {Tanaka} K. E.~I.,  {Ramsey} J.~P.,  {Zhang} Y.,   {Tan} J.~C.,  2023, \mn@doi [\apj] {10.3847/1538-4357/acbd47}, \href {https://ui.adsabs.harvard.edu/abs/2023ApJ...947...40S} {947, 40}

\bibitem[\protect\citeauthoryear{{Tan}, {Beltr{\'a}n}, {Caselli}, {Fontani}, {Fuente}, {Krumholz}, {McKee}  \& {Stolte}}{{Tan} et~al.}{2014}]{Tan2014prpl.conf..149T}
{Tan} J.~C.,  {Beltr{\'a}n} M.~T.,  {Caselli} P.,  {Fontani} F.,  {Fuente} A.,  {Krumholz} M.~R.,  {McKee} C.~F.,   {Stolte} A.,  2014, in {Beuther} H.,  {Klessen} R.~S.,  {Dullemond} C.~P.,   {Henning} T.,  eds, Protostars and Planets VI. pp 149--172 (\mn@eprint {arXiv} {1402.0919}), \mn@doi{10.2458/azu_uapress_9780816531240-ch007}

\bibitem[\protect\citeauthoryear{{Tanaka}, {Tan}  \& {Zhang}}{{Tanaka} et~al.}{2017}]{Tanaka2017ApJ...835...32T}
{Tanaka} K. E.~I.,  {Tan} J.~C.,   {Zhang} Y.,  2017, \mn@doi [\apj] {10.3847/1538-4357/835/1/32}, \href {https://ui.adsabs.harvard.edu/abs/2017ApJ...835...32T} {835, 32}

\bibitem[\protect\citeauthoryear{{Tanikawa}, {Tajitsu}, {Honda}, {Maehara}, {Sato}, {Masuda}, {Omiya}  \& {Izumiura}}{{Tanikawa} et~al.}{2026}]{Tanikawa2026arXiv260121125T}
{Tanikawa} A.,  {Tajitsu} A.,  {Honda} S.,  {Maehara} H.,  {Sato} B.,  {Masuda} K.,  {Omiya} M.,   {Izumiura} H.,  2026, \mn@doi [arXiv e-prints] {10.48550/arXiv.2601.21125}, \href {https://ui.adsabs.harvard.edu/abs/2026arXiv260121125T} {p. arXiv:2601.21125}

\bibitem[\protect\citeauthoryear{{Telkamp} et~al.,}{{Telkamp} et~al.}{2025}]{Telkamp2025ApJ...986...15T}
{Telkamp} Z.,  et~al., 2025, \mn@doi [\apj] {10.3847/1538-4357/adcd79}, \href {https://ui.adsabs.harvard.edu/abs/2025ApJ...986...15T} {986, 15}

\bibitem[\protect\citeauthoryear{{Tokovinin}}{{Tokovinin}}{2026}]{Tokovinin2026arXiv260105006T}
{Tokovinin} A.,  2026, \mn@doi [arXiv e-prints] {10.48550/arXiv.2601.05006}, \href {https://ui.adsabs.harvard.edu/abs/2026arXiv260105006T} {p. arXiv:2601.05006}

\bibitem[\protect\citeauthoryear{{Tokovinin} \& {Moe}}{{Tokovinin} \& {Moe}}{2020}]{TokovininMoe2020MNRAS.491.5158T}
{Tokovinin} A.,  {Moe} M.,  2020, \mn@doi [\mnras] {10.1093/mnras/stz3299}, \href {https://ui.adsabs.harvard.edu/abs/2020MNRAS.491.5158T} {491, 5158}

\bibitem[\protect\citeauthoryear{{Tramper}, {Sana}, {de Koter}  \& {Pauwels}}{{Tramper} et~al.}{2025}]{Tramper2025arXiv250918431T}
{Tramper} F.,  {Sana} H.,  {de Koter} A.,   {Pauwels} T.,  2025, \mn@doi [arXiv e-prints] {10.48550/arXiv.2509.18431}, \href {https://ui.adsabs.harvard.edu/abs/2025arXiv250918431T} {p. arXiv:2509.18431}

\bibitem[\protect\citeauthoryear{{Vasyunina}}{{Vasyunina}}{2010}]{Vasyunina2010PhDT........23V}
{Vasyunina} T.,  2010, PhD thesis, Ruprecht-Karls University of Heidelberg, Germany

\bibitem[\protect\citeauthoryear{{Wall}, {McMillan}, {Mac Low}, {Klessen}  \& {Portegies Zwart}}{{Wall} et~al.}{2019}]{Wall2019ApJ...887...62W}
{Wall} J.~E.,  {McMillan} S. L.~W.,  {Mac Low} M.-M.,  {Klessen} R.~S.,   {Portegies Zwart} S.,  2019, \mn@doi [\apj] {10.3847/1538-4357/ab4db1}, \href {https://ui.adsabs.harvard.edu/abs/2019ApJ...887...62W} {887, 62}

\bibitem[\protect\citeauthoryear{{Wang}, {Li}, {Abel}  \& {Nakamura}}{{Wang} et~al.}{2010}]{Wang2010ApJ...709...27W}
{Wang} P.,  {Li} Z.-Y.,  {Abel} T.,   {Nakamura} F.,  2010, \mn@doi [\apj] {10.1088/0004-637X/709/1/27}, \href {https://ui.adsabs.harvard.edu/abs/2010ApJ...709...27W} {709, 27}

\bibitem[\protect\citeauthoryear{{Wilking}, {Gagn{\'e}}  \& {Allen}}{{Wilking} et~al.}{2008}]{Wilking2008hsf2.book..351W}
{Wilking} B.~A.,  {Gagn{\'e}} M.,   {Allen} L.~E.,  2008, in {Reipurth} B.,  ed., , Vol.~5, Handbook of Star Forming Regions, Volume II.
Astronomical Society of the Pacific, p.~351, \mn@doi{10.48550/arXiv.0811.0005}

\bibitem[\protect\citeauthoryear{{Wu}, {Tan}, {Christie}, {Nakamura}, {Van Loo}  \& {Collins}}{{Wu} et~al.}{2017}]{Wu2017ApJ...841...88W}
{Wu} B.,  {Tan} J.~C.,  {Christie} D.,  {Nakamura} F.,  {Van Loo} S.,   {Collins} D.,  2017, \mn@doi [\apj] {10.3847/1538-4357/aa6ffa}, \href {https://ui.adsabs.harvard.edu/abs/2017ApJ...841...88W} {841, 88}

\bibitem[\protect\citeauthoryear{{Zhang} \& {Tan}}{{Zhang} \& {Tan}}{2018}]{ZhangTan2018ApJ...853...18Z}
{Zhang} Y.,  {Tan} J.~C.,  2018, \mn@doi [\apj] {10.3847/1538-4357/aaa24a}, \href {https://ui.adsabs.harvard.edu/abs/2018ApJ...853...18Z} {853, 18}

\bibitem[\protect\citeauthoryear{{Zummer}, {Harmanec}, {Barlow}, {Blackford}  \& {{\v{S}}vr{\v{c}}kov{\'a}}}{{Zummer} et~al.}{2025}]{Zummer2025arXiv251115544Z}
{Zummer} M.,  {Harmanec} P.,  {Barlow} B.,  {Blackford} M.,   {{\v{S}}vr{\v{c}}kov{\'a}} J.,  2025, \mn@doi [arXiv e-prints] {10.48550/arXiv.2511.15544}, \href {https://ui.adsabs.harvard.edu/abs/2025arXiv251115544Z} {p. arXiv:2511.15544}

\makeatother
\end{thebibliography}

\bsp	
\label{lastpage}
\end{document}